\documentclass[epj]{svjour}
\usepackage{latexsym}
\usepackage{amssymb,amsmath}
\usepackage{graphicx}
\usepackage[subrefformat=parens,labelformat=parens,caption=false]{subfig}
\usepackage{color}
\usepackage[english]{babel}
\usepackage{makecell}

\begin{document}
\title{Emergent collective dynamics of\\
bottom-heavy squirmers under gravity}

\author{Felix R\"uhle\inst{1} \and Holger Stark\inst{1}
}                     

\institute{Technische Universit\"at Berlin, Institut f\"ur Theoretische Physik, Hardenbergstr. 36, D-10623 Berlin, Germany}
\abstract{We present the results of hydrodynamic simulations using the method of multi-particle collision dynamics for a system of squirmer microswimmers moving under the influence of 
gravity at low Reynolds numbers. In addition, the squirmers are bottom-heavy so that they experience a torque which aligns them along the vertical. The 
squirmers interact hydrodynamically by the flow fields of a stokeslet and rotlet, which are initiated by the acting gravitational force 
and torque, respectively, and by their own flow fields.
By varying the ratio of swimming to bulk sedimentation velocity and the torque, we determine state diagrams for the emergent collective dynamics of neutral squirmers as well as strong pushers and pullers. For low swimming velocity and torque we observe conventional sedimentation, while the sedimentation profile becomes inverted when their values are increased. For neutral squirmers we discover convective rolls of circulating squirmers  between both sedimentation states, which sit at the bottom of the 
system and are fed by plumes made of collectively sinking squirmers. At larger torques porous clusters occur that spawn single squirmers. The two latter states can also occur transiently starting from a uniform squirmer distribution and then disappear in the 
long-time limit. For strong pushers and pullers only weak plume formation is observed.
 %
\PACS{
      {47.55.P}{Buoyancy-driven flows; convection}   \and
      {47.57.ef} {Sedimentation and migration} \and
      {47.63.Gd}{Swimming microorganisms} \and
      {87.18.Hf}{Spatiotemporal pattern formation in cellular populations}
     } 
} 
\maketitle
\section{Introduction}
\label{sec:intro}

Active entities consume energy locally in order to self-propel without an external force. 
When these non-equi\-lib\-rium objects move collectively, fascinating patterns emerge both on the
macroscopic and microscopic scale~\cite{Ramaswamy2010,RomanczukSchimansky-Geier2012,ElgetiGompper2015,BechingerVolpe2016,ZoettlStark2016}, 
such as flocks of birds~\cite{CavagnaViale2010}, motility-induced phase separation~\cite{FilyMarchetti2012,ButtinoniSpeck2013,RednerBaskaran2013,CatesTailleur2015}, swarming~\cite{ThutupalliHerminghaus2011,CopelandWeibel2009,OyamaYamamoto2016,JeckelDrescher2019}, and active turbulence~\cite{WensinkYeomans2012,DunkelGoldstein2013}. Of particular interest are microswimmers, \textit{i.e.}, organisms or synthetic particles that self-propel in a fluid at low Reynolds numbers~\cite{Purcell1977,MarchettiSimha2013}. At higher densities, hydrodynamic interactions
between the microswimmers become important and influence their collective dynamics~\cite{IshikawaLocseiPedley2008,IshikawaPedley2008,MolinaYamamoto2013,BlaschkeStark2016,TheersGompper2018}. 

Studying microswimmers under gravity is important because often they are not neutrally buoyant~\cite{PalacciBocquet2010,NashCates2010,Stark2016,ShenLintuvuori2019}. In such a setting non-equilibrium sedimentation has been observed~\cite{PalacciBocquet2010,JungValles2014,GinotCottin-Bizonne2015}
accompanied by polar order along the vertical \cite{EnculescuStark2011} and convection~\cite{KuhrStark2017}. Numerical hydrodynamic studies also discovered two-dimensional Wigner fluids
and swarming under strong gravity~\cite{KuhrStark2019}, as well as fluid pumps in a parabolic potential \cite{HennesStark2014}.
In experimental systems swimming under the influence of external fields generates intriguing and surprising phenomena such as the formation of thin layers of motile phytoplankton in coastal regions~\cite{DurhamStocker2009}, algae in bound dancing states~\cite{DrescherGoldstein2009}, and hovering rafts of  active emulsion droplets~\cite{KruegerMaass2016}.

Many microswimmers also perform \textit{gravitaxis}, which is the ability to align (anti-)parallel to the direction of gravity. The 
gravitational torque to achieve this alignment can either result from hydrodynamic drag of a microswimmer 
with asymmetric morphology~\cite{Roberts2006,tenHagenBechinger2014,SenguptaStocker2017} or from bottom heaviness, 
\textit{i.e.}, when the center of mass is offset relative to the geometrical center~\cite{Kessler1985,DurhamStocker2009,CampbellEbbens2013,WolffStark2013}. Gravitactic swimmers can induce 
an overturning instability when accumulating with higher density at the top boundary (reminiscent of the Rayleigh-Taylor instability), which then initiates various patterns of bioconvection \cite{PlessetWinet1974,ChildressSpiegel1975,NewellWhitehead1969,PedleyKessler1992}. 
However, also \textit{gyrotaxis} clearly plays an important role in such settings~\cite{PedleyKessler1992,HillKessler1989,BeesHill1997,GhoraiHill1999,CzirokKessler2000,DesaiArdekani2017}.
There, the collective dynamics of microswimmers depends on the combined action of gravity and hydrodynamic flow \cite{Kessler1985,PedleyKessler1992}.
The involvement of physiological aspects in biotic pattern formation has also been discussed~\cite{MachemerTakahashi1991,OoyaBaba1992,Roberts2010}.

In theory microswimmer systems under gravity have been investigated in the past using the versatile spherical squirmer model swimmer \cite{ShenLintuvuori2019,KuhrStark2017,KuhrStark2019,RuehleStark2018,BrumleyPedley2019,FaddaYamamoto2020}. 
Here, we simulate around 900 bottom-heavy squirmers under gravity with full hydrodynamics using the method of 
multi-particle collision dynamics (MPCD) \cite{MalevanetsKapral1999,NoguchiGompper2007}. 
Varying the ratio of swimming to bulk sedimentation velocity and 
the gravitational torque due to bottom heaviness, we determine state diagrams for neutral as well as strong
pusher and puller squirmers. While for low swimming velocity and torque conventional sedimentation is recovered, the sedimentation profile becomes inverted when increasing their values. For neutral squirmers we discover a rich phenomenology between both states including a state where plumes consisting of collectively sinking squirmers feed convective rolls at the bottom of the system and dense clusters which spawn single squirmers. These two states can also occur transiently starting from a uniform squirmer distribution and then disappear in the long-time limit. For strong pushers and pullers only weak plume formation is observed. We 
thoroughly characterize all states by different quantities.

In the following, in sect.\ \ref{sec:methods} we introduce the squirmer model swimmer and the simulation method
of multi-par\-ti\-cle collision dynamics. Then, in sect.\ \ref{sec:results} we present all our results. We start with the state diagram of neutral squirmers followed by a detailed characterization of the different states and also look at strong pushers and pullers. Finally, we finish with a summary and conclusions.

\section{Squirmer model swimmer and simulation method}
\label{sec:methods}

\subsection{Spherical squirmer}
\label{sec:squirmer}

Swimming on the micron scale is dominated by friction \cite{Taylor1951,Purcell1977}. Hence, hydrodynamics is captured by the Stokes equations:
\begin{eqnarray}
\nabla \cdot \mathbf{u} & = & 0 \\
\eta \nabla^2 \mathbf{u} & = & \nabla p \, ,
\end{eqnarray}
where $\mathbf{u}$ and $p$ are the fluid velocity and pressure fields, respectively.

Biological microswimmers often propel themselves by collective beating patterns of cilia, which create flow fields along their cell surfaces \cite{ElgetiGompper2015,LaugaPowers2009}. Also artificial microswimmers exist that either use phoretic self-propulsion mechanisms, such as diffusiophoresis and thermophoresis in the case of active colloids \cite{ElgetiGompper2015,PalacciChaikin2013,ButtinoniBechinger2012}, or Marangoni stresses in the case of active  emulsion droplets \cite{ThutupalliHerminghaus2011} in order to generate such surface flow fields.
A simple and effective approximation to all these swimming mechanisms is offered by the spherical squirmer model \cite{Lighthill1952,Blake1971}, where an axisymmetric tangential flow field on the surface is prescribed: 
\begin{equation}
\label{eq:surface_field}
\mathbf{u}(\mathbf{r})\vert_{r=R} = \sum_{n=1}^{\infty}B_n\dfrac{2 P_n^\prime(\mathbf{e}\cdot\mathbf{\hat{r}})}{n(n+1)}\left[
-\mathbf{e} + (\mathbf{e}\cdot \mathbf{\hat{r}})\mathbf{\hat{r}}\right]\, .
\end{equation}
Here, $\mathbf{e}$ is the swimmer orientation vector, $R$ is the swimmer radius, $P_n$ is the $n$th Legendre polynomial,
and $P_n^\prime$ means its first derivative.

Typically, the expansion is truncated after the second term, leaving the two relevant modes $B_1$ and $B_2$. Then, the flow field generated by the surface field of eq.\ (\ref{eq:surface_field})
in the surrounding fluid is \cite{Blake1971,PakLauga2014}
\begin{align}
\label{eq:squirmer_field}
	\begin{split}
		\mathbf{u}(\mathbf{r}) = &  \frac{B_1}{2}\Biggl[\left(-\frac{R}{r}\left[\mathbf{e}+(\mathbf{e}\cdot\mathbf{\hat{r}})\mathbf{\hat r}\right]+\frac{R^3}{r^3}\left[-\mathbf{e}+3(\mathbf{e}\cdot\mathbf{\hat{r}})\mathbf{\hat{r}}\right]\right) \\
		&  - \beta\frac{R^2}{r^2}\left(-\mathbf{\hat{r}}+3\left(\mathbf{e}\cdot\mathbf{\hat{r}}\right)^2\mathbf{\hat{r}}\right) + \mathcal{O}\left(\frac{R^4}{r^4}\right)\Biggr] \, 
	\end{split}
\end{align}
where $\beta=B_2/B_1$ is the squirmer-type parameter.

\subsubsection{Free squirmer}
The squirmer induces a  hydrodynamic source dipole and for $\beta \neq 0$ also a force di\-pole,
the far fields of which decay as $1/r^3$ and $1/r^2$, respectively. Swimmers with $\beta = 0$ are called neutral squirmers, 
while $\beta > 0$ generates pullers and $\beta < 0$ pushers. Since free squirmers are force-free, a stokeslet term with a 
flow field decaying as $1/r$ is not allowed but apears in eq.~\eqref{eq:squirmer_field}~\cite{Lighthill1952,Blake1971}. 
The reason is that for a moving squirmer also the swimming velocity $v_0\mathbf{e}$ contributes to its surface velocity field, which is not included in eq.\ (\ref{eq:surface_field}). 
Thus, following Pak and Lauga~\cite{PakLauga2014}
the flow field of eq.\ \eqref{eq:squirmer_field} has to be interpreted as the pumping field generated by a squirmer held at a constant position by a force. 
This stalling force $\mathbf{F}_a$ is given by the balance equation \cite{PakLauga2014}:
\begin{equation}
\label{eq:force_balance}
\mathbf{F}_a - 6\pi\eta R\mathbf{v} = 0 \, ,
\end{equation}
where here $\mathbf{v}$ is the swimming velocity of the freely moving squirmer. One can calculate the stalling force using Lamb's solution to the Stokes equations~\cite{Lamb1932,KimKarrila2013,PakLauga2014},
$\mathbf{F}_a = 4\pi \eta R \nabla(\mathbf{e} \cdot \mathbf{r})B_1
= 4\pi \eta R B_1 \mathbf{e}$,
and arrive at the known relation $\mathbf{v} = \dfrac{2}{3}B_1\mathbf{e}$ with the swimming speed $v_0:=\frac{2}{3}B_1$.

In a freely translating squirmer, the Stokes flow field initiated by the pumping force
is no longer present. Thus, in eq.~\eqref{eq:squirmer_field}
the stokeslet vanishes and the source-dipole term is modified leading to the flow field of a free squirmer \cite{Blake1971,PakLauga2014},
\begin{align}
\label{eq:squirmer_free_field}
	\begin{split}
		\mathbf{u}_\mathrm{free}(\mathbf{r}) = &  B_1\Biggl[\frac{1}{3}\frac{R^3}{r^3}\left[-\mathbf{e}+3(\mathbf{e}\cdot\mathbf{\hat{r}})\mathbf{\hat{r}}\right] \\
		&  - \frac{\beta}{2}\frac{R^2}{r^2}\left(-\mathbf{\hat{r}}+3\left(\mathbf{e}\cdot\mathbf{\hat{r}}\right)^2\mathbf{\hat{r}}\right) + \mathcal{O}\left(\frac{R^4}{r^4}\right)\Biggr] \, .
	\end{split}
\end{align}

\subsubsection{Squirmer under gravity}
Adding the gravitational force $-mg\mathbf{e}_z$ 
modifies the force balance of eq.\ (\ref{eq:force_balance}) and yields for the total squirmer velocity,
\begin{equation}
\label{eq:velocity_gravity}
\mathbf{v} = v_0\mathbf{e} - mg/(6\pi\eta R)\mathbf{e}_z,
\end{equation} 

As in our previous publications~\cite{KuhrStark2017,RuehleStark2018} we introduce the velocity ratio 
$\alpha: = v_0/v_\mathrm{sed}$ to compare 
the self-propul\-sion to the bulk sedimentation velocity, $v_\mathrm{sed} = {mg}/{6\pi\eta R}$.

The gravitational force adds a Stokes flow field to the free-squirmer solution of eq. (\ref{eq:squirmer_free_field}), which as usual
contains a stokeslet and a source-dipole contribution:
\begin{align}
\label{eq:grav_flow_st}
\mathbf{u}_\mathrm{st}^g  & =-\frac{3}{4}v_\mathrm{sed}\dfrac{R}{r} \left(\mathbf {e}_z + 
\,\frac{z}{r}\,
\mathbf{\hat{r}}\right)\\
\label{eq:grav_flow_sd}
\mathbf{u}_\mathrm{sd}^g & = \frac{1}{4}v_\mathrm{sed}\dfrac{R^3}{r^3}\left(-\mathbf{e}_z + 3
\,\frac{z}{r}\,
\mathbf{\hat{r}}\right) \, ,
\end{align}
where we introduced the coordinate along the vertical $z = \mathbf{r} \cdot  \mathbf{e}_z$.
Due to their long-range nature it is important to always take stokeslet flow fields into account when they occur. This has been shown in experimental studies of the dancing motion of Volvox algae \cite{DrescherGoldstein2009} 
or the Stokesian dynamics of swimmers in a harmonic trap \cite{HennesStark2014}.

\subsubsection{Squirmer with bottom heaviness}
In this article we assume the spherical squirmer to be bottom-heavy, \textit{i.e.}, its center of mass
has an offset $r_0$ from the geometrical center~\cite{WolffStark2013} such that a torque $mgr_0(-\mathbf{e}_z\times\mathbf{e})$ 
acts on the swimmer. Balancing external torque and rotational friction torque $- 8\pi \eta R^3 \mathbf{\Omega}$, we find 
the angular velocity
\begin{equation}
\label{eq:angular_velocity_bh}
\boldsymbol{\Omega} = \frac{3}{4}\frac{v_0}{R}\frac{r_0}{R\alpha}(-\mathbf{e}_z\times\mathbf{e}) \, .
\end{equation} 
We will later use the dimensionless parameter
\begin{equation}
\frac{r_0}{R\alpha} =
\frac{R}{v_0}  \, \frac{mg r_0}{6 \pi \eta R^3}
\end{equation} 
to quantify the strength of the external torque. It compares - up to the factor 3/4
- the characteristic time scale of  self-propulsion, 
$R/v_0$, to the characteristic time of reorientation by bottom heavi\-ness, 
$8\pi \eta R^3 / (mgr_0)$.
Sometimes it is called the gyrotactic orientation parameter \cite{PedleyKessler1992,PedleyKessler1987}.

The rotating squirmer generates the flow field of a rotlet \cite{KimKarrila2013},
\begin{equation}
\label{eq:grav_flow_rot}
\mathbf{u}_r^\mathrm{bh} = 
\frac{R^3}{r^2} \boldsymbol{\Omega} \times \mathbf{\hat{r}} = \frac{3}{4}v_0 \frac{r_0}{R\alpha} \frac{R^2}{r^2} \left((\mathbf{e}\cdot\mathbf{\hat{r}}) \mathbf{e}_z - 
\,\frac{z}{r}\, \mathbf{e}\right).
\end{equation}
Note that the spatial decay of this flow field is the same as for pusher and puller squirmers but is more long-ranged compared to the neutral squirmer. However, $\mathbf{u}_r^\mathrm{bh}$ vanishes when the squirmer is aligned with the vertical, meaning $\mathbf{e} = \mathbf{e}_z$.

\subsection{Multi-particle collision dynamics}
\label{sec:mpcd}
In the following we present the algorithm for performing  simulations with multi-particle collision dynamics
that we have already used in the past \cite{BlaschkeStark2016,KuhrStark2017,KuhrStark2019}. Therefore, we only summarize it here. The algorithm is implemented in a massively parallelized code, which runs on 
a computer cluster. In addition, we also provide the parameters used in our simulations.

\subsubsection{Algorithm}
We numerically solve the Navier-Stokes equations at low Reynolds number using the mesoscale method  of multi-particle collision dynamics (MPCD) \cite{MalevanetsKapral1999,PaddingLouis2006,NoguchiGompper2007,ZoettlStark2018}. Thermal noise is automatically included in this particle-based solver. 
Since the Reynolds numbers employed in our simulations are  smaller than one, we effectively obtain solutions of the Stokes equations where any inertia is neglected.

In the MPCD method the fluid is composed of point particles of mass $m_0$ that are kept at temperature $T_0$. A simulation step consists of the fluid particles performing  consecutive streaming and collision steps. 
During the streaming step, which has a duration $\Delta t$, each fluid particle $i$ moves ballistically with its velocity $\mathbf{v}_i$ according to $\mathbf{r}_i(t+\Delta t) = \mathbf{r}_i(t)+\mathbf{v}_i\Delta t$ (see footnote
\footnote{Our system height is smaller than the sedimentation length of the fluid so that gravity 
acting on the fluid particles can be neglected.}).
The duration $\Delta t$ is a simulation parameter that controls the fluid viscosity~\cite{PaddingLouis2006,NoguchiGompper2008}. 
During the streaming step fluid momentum is advected and also transferred to swimmers or absorbed by walls. Furthermore, 
boundary conditions need to be applied to the squirmer surfaces and to bounding walls. For this we employ the bounce-back rule~\cite{PaddingLouis2005} to implement either the surface slip velocity field of a squirmer from eq.~\eqref{eq:surface_field} or the
no-slip boundary condition for walls. The dynamics of the  squirmers themselves is also computed during the streaming step. 
We perform 20 molecular dynamics steps during each streaming step using Velocity-Verlet integration together with 
the gravitational force and steric interactions between squirmers \cite{ZoettlStark2018}.

The purpose of the collision step is to exchange momentum between fluid particles. To that end, the simulation box is divided into cubical cells of edge length $a_0$.  We also use this length to define the fluid particle density as the average particle number $n_\mathrm{fl}$  per collision cell. Within each cell fluid-particle velocities are updated with the help of a collision operator, for which we use the MPC-AT+a rule \cite{NoguchiGompper2007,ZoettlStark2018}. Thus, a thermostat is set up and both linear and angular momentums are conserved~\cite{NoguchiGompper2007}. We also apply a 
grid shift for each new collision step in order to enforce Galilean invariance \cite{IhleKroll2003}. 
During the collision step fluid and squirmers/bounding walls interact as well: 
if a collision cell overlaps with a boundary or a squirmer, the overlapping volume is filled with virtual fluid particles to ensure the fluid density $n_\mathrm{fl}$ remains the same~\cite{LamuraKroll2001,ZoettlStark2018}. After the collision step the momentum gain of the virtual fluid particles is transferred to the involved squirmer where the virtual particles are located.

The flow fields calculated with the MPCD method are accurate on length scales larger than the mean free path of the fluid particles. Using large enough squirmer radii, the hydrodynamic behaviour of squirmer microswimmers is therefore well reproduced by the MPCD method~\cite{DowntonStark2009,GoetzeGompper2010}. Thus it has widely been used to simulate a variety of settings~\cite{GoetzeGompper2010,ZoettlStark2014,BlaschkeStark2016,KuhrStark2017,RuehleStark2018,TheersWinkler2016Soft}.

\begin{table*}[ht]
	\centering
	\caption{Characteristics of neutral squirmer states}
	\label{tab:states}
	\begin{tabular}{l|l|l|l}
		state & \makecell[l]{sedimentation profile} & \makecell[l]{sinking clusters/plumes} & convection \\ 
		& $\qquad $figs.\ \ref{fig:density_profiles}, \ref{fig:g-005_density_profiles} & 
		$\qquad$figs.\ \ref{fig:clusters}, \ref{fig:mean_sinking} & \makecell[l]{$\enspace$figs.\ \ref{fig:current_density},\ref{fig:larger_current},\ref{fig:spawning_2dhist}}  \\
		\hline\hline
		sedimentation & \makecell[l]{exponential decay}& \makecell[l]{rare} & weak \\ \hline
		\makecell[l]{inverted sedimentation}& \makecell[l]{exponential increase}& \makecell[l]{very rare} & none \\ \hline
		\makecell[l]{plumes and \\ convective rolls} & \makecell[l]{pronounced maximum \\ at the bottom} & \makecell[l]{pronounced} & strong \\\hline
		\makecell[l]{spawning cluster} & \makecell[l]{pronounced maximum \\ at the bottom \\ \hline strong depletion \\ in the middle} & \makecell[l]{very rare} & none
	\end{tabular}
\end{table*}

\subsubsection{Parameters}
For the most part, we use the parameters presented below, in case of deviations we have stated them in the text.
We use the duration $\Delta t = 0.02a_0\sqrt{m_0/k_BT_0}$ for the streaming step and a fluid particle density of $n_\mathrm{fl} = 10$. This implies a viscosity of $\eta = 16.05 \sqrt{m_0k_BT_0}/a_0^2$~\cite{NoguchiGompper2008,ZoettlPHD}. For our squirmers we use a radius $R=4a_0$, therefore  the translational and rotational thermal diffusivities in bulk fluid are
$D_T = k_BT /(6\pi\eta R) \approx 8\cdot 10^{-4} a_0\sqrt{k_BT_0/m_0}$  and
$D_R = k_BT /(8\pi\eta R^3) \approx 4\cdot 10^{-5}\sqrt{k_BT_0/m_0}/a_0$, respectively. We choose $B_1=0.1\sqrt{k_BT_0/m_0}$ and thus have the active P\'{e}clet number $\mathrm{Pe} = Rv_0 / D_T = 330$, which is comparable in order of magnitude to bacterial swimmers, such as \emph{E. Coli}\ \cite{ChattopadhyayWu2006} or \emph{B. subtilis}\ \cite{JanosiHorvath1998},
with P\'{e}clet numbers around 100.
Furthermore, the ballistic time scale of the squirmer is $R/v_0 =3000 \Delta t$.
The Reynolds number $\mathrm{Re} = v_0Rn_\mathrm{fl}/\eta = 0.17$ is relatively high for creeping flow systems. Lowering it further would mean considerably additional computational cost. However, viscous effects are definitely dominant in our simulations.

Typically, the volume density of squirmers was held constant at $10\%$, simulating 914 squirmers in a box with a system size of 
$108a_0 \times 108a_0 \times 210a_0$, where the latter length is the box height. This means that in a cross-sectional slab of width $2R$ perpendicular to the direction of gravity, the mean area fraction of squirmers is 15\%. We use no-slip boundary conditions at the top and bottom walls  and periodic boundary conditions in the horizontal plane. As initial condition we always choose uniformly 
distributed squirmers such that no squirmers overlap.

For our study of bottom-heavy squirmers we vary the velocity ratio $\alpha$ by  changing the gravitational acceleration $g$, 
which in an experiment depends on the density mismatch between swimmer and fluid. We find that increasing $\alpha$ from zero up to around 7 captures the relevant features in our squirmer simulations. Values in this range can be obtained experimentally for example with Volvox algae in water~\cite{DrescherTuval2010} or active emulsion droplets in a mixture of $\mathrm{H_2O}$ and $\mathrm{D_2O}$~\cite{KruegerMaass2016}. The rescaled torque $r_0/R\alpha$ is 
varied by changing both $g$ and the center-of-mass offset $r_0$. Note that the torque depends on real mass rather than buoyant 
mass. In order to account for  this difference, in ref.~\cite{WolffStark2013} the parameter $r_0$ was redefined as $r_0m/\Delta m$, 
with the buoyant mass $\Delta m$.  This way we can use the same parameter $\alpha$ in both the torque and force equations. Alternatively, one could also assume that $\Delta m \approx m$~\cite{WolffStark2013}.

To analyze the data of our simulation runs, we save the position, orientation, as well as translational and angular velocities for each squirmer every 1000th time step of duration $\Delta t$.

Note for most densely packed squirmers a depletion of the MPCD fluid particles is observed between the squirmers due to finite compressibility \cite{TheersGompper2018}. In the states discussed in sects.~\ref{sec:convective_roll} and \ref{sec:spawning} the squirmers are not most densely packed. Nevertheless, we ran one simulation for each of the states at higher fluid particle density $n_\mathrm{fl}=80$ and decreased self-propulsion velocity $B_1=0.01$. This leaves the P\'{e}clet number almost constant~\cite{TheersGompper2018} while lowering compressibility. In these simulations we could confirm that the phenomenology stays the same as for the lower fluid particle density, which we typically use.

\section{Results}      
\label{sec:results}

\begin{figure*}[h!]
\centering
\includegraphics[width=0.95\textwidth]{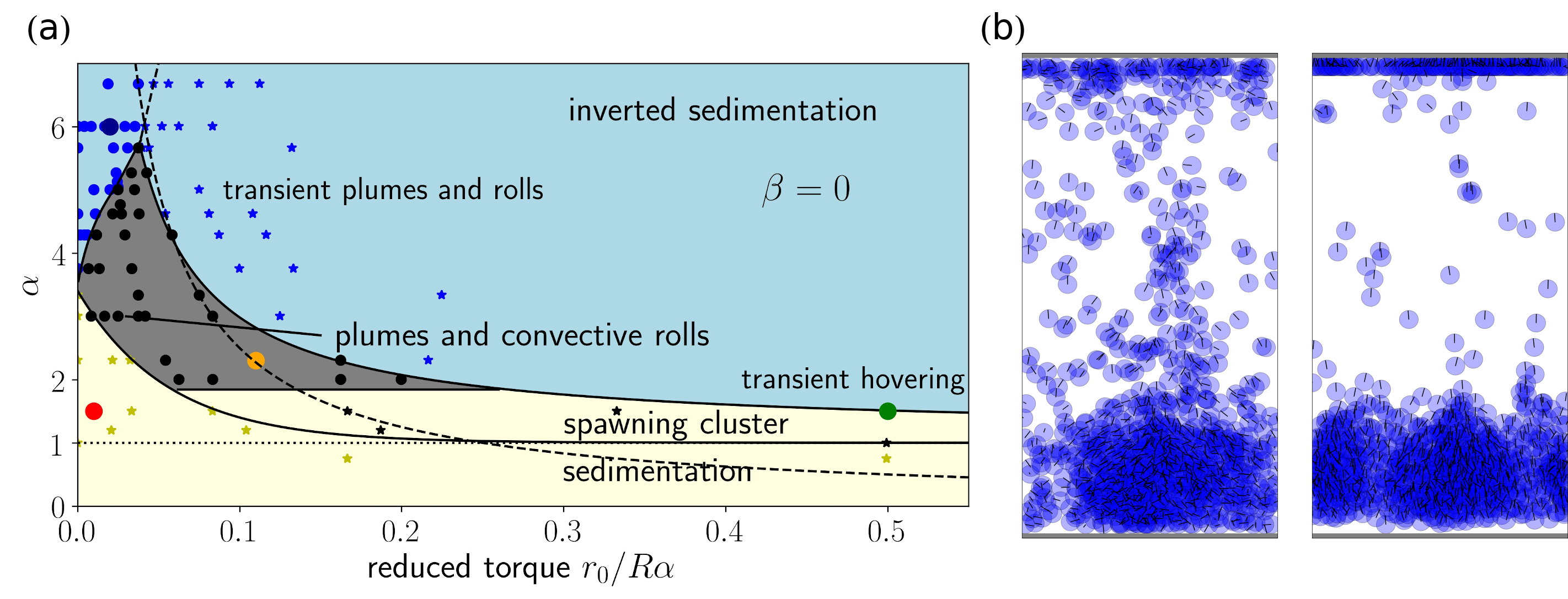}

\caption{(a) Observed states of neutral squirmers in the parameter space $\alpha$ versus $r_0/R\alpha$. All symbols represent simulated systems. The colored dots mark states for which we provide videos M1-M4  in the supplemental video, density profiles in fig.\ \ref{fig:density_profiles}, and for two of them snaphots on the right. The dashed line shows a
hyperbola, which follows from eq.\ (\ref{eq.estimate}), where stokeslet vorticity and rotation due to bottom heaviness are balanced.
(b) Snapshots of a plume and convection roll for $\alpha=2.3$ and $r_0/R\alpha=0.11$ (left) and of a spawning cluster for $\alpha=1.5$
and $r_0/(R\alpha)=0.5$ (right).
}
\label{fig:neutral_states}
\end{figure*}
In our hydrodynamic simulations we explored the dynamics of bottom-heavy squirmers under gravity.  Depending on 
the velocity ratio $\alpha: = v_0/v_\mathrm{sed}$ and the strength of the gravitational torque, we observed a variety of stable and transient states.  In fig.~\ref{fig:neutral_states}(a) we show the state diagram in the parameter space $\alpha$ versus reduced gravitational torque $r_0/R\alpha$, which we determined in simulations with neutral squirmers.
We have marked four exemplary states by colored points, for which we show the density profiles in fig.~\ref{fig:density_profiles} in the same colors and videos M1-M4  in the supplemental material. In the following, we introduce the main phenomenology  of the observed states.

\begin{figure}
\includegraphics[width=0.45\textwidth]{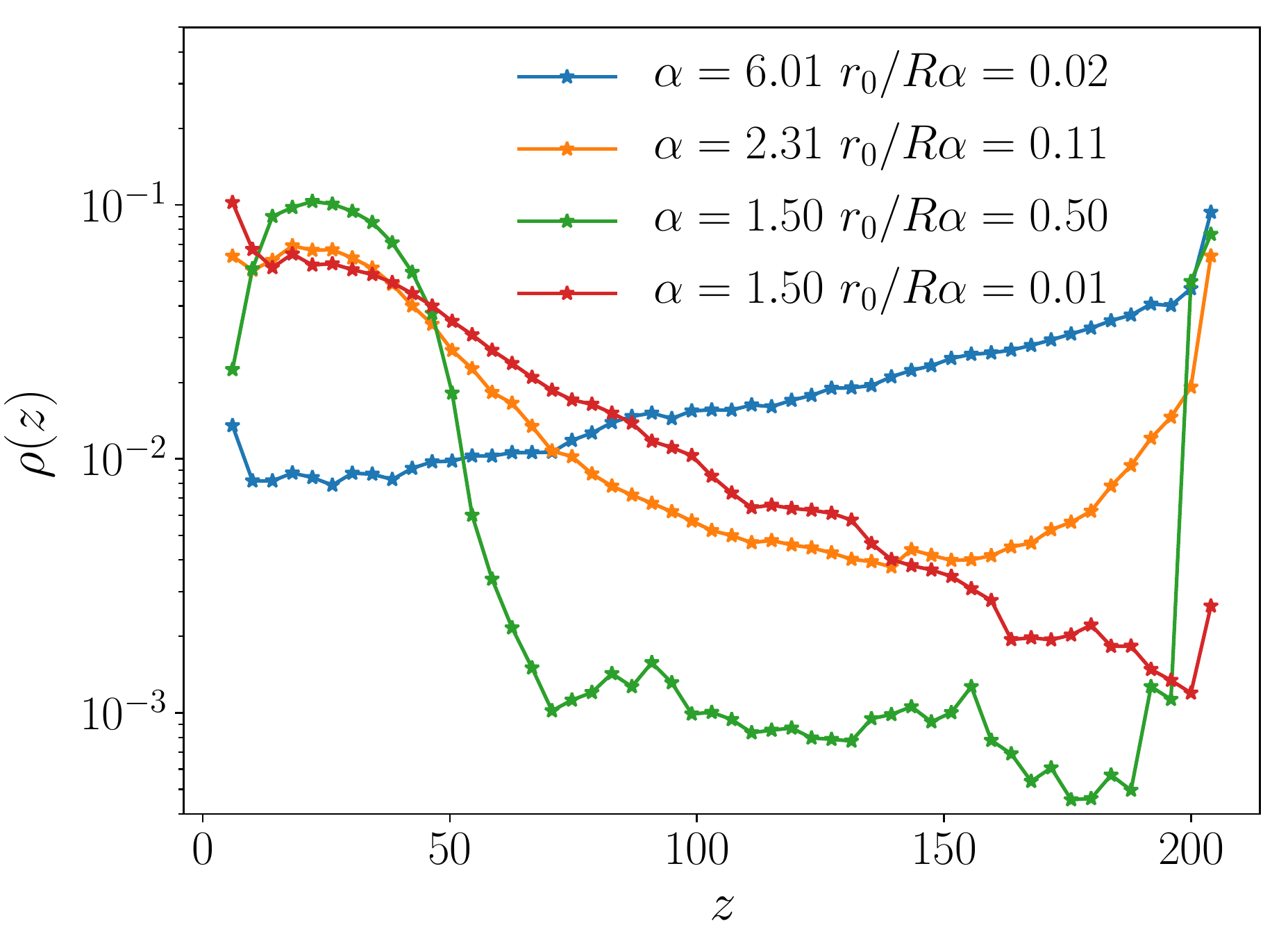}
\caption{Density profiles for collective squirmer states marked by dots of the same color in the state diagram of fig.~\ref{fig:neutral_states}(a). 
The profiles belong to conventional non-equilibrium sedimentation  (red), inverted sedimentation (blue), stable plumes and convection rolls (orange), and the spawning cluster (green).}
\label{fig:density_profiles}
\end{figure}

The horizontal dotted line at $\alpha=1$ in fig.~\ref{fig:neutral_states}(a) marks the upper limit, 
below which isolated squirmers sink down in a bulk fluid and settle at a finite distance from a lower bounding wall~\cite{RuehleStark2018}.
Hydrodynamically interacting squirmers show a non-equilibrium sedimentation state, which persists to $\alpha \approx 3$ for weak torques.
For zero torque the non-equilibrium sedimentation was already observed in ref.\ \cite{KuhrStark2017}.
A typical sedimentation profile at $\alpha=1.5$ and $r_0/R\alpha=0.01$  is depicted in fig.~\ref{fig:density_profiles}. 
In contrast, at large $\alpha$ and torques, the orientational bias of the squirmers leads to their enrichment at the top wall and
inverted sedimentation occurs, which is even observable for small torques.
The corresponding inverted sedimentation profile for $\alpha=6.01$ and $r_0/R\alpha = 0.02$ is  shown in 
fig.~\ref{fig:density_profiles}. For dilute systems of bottom-heavy active particles such states were already  described in ref.\ \cite{WolffStark2013}. Between sedimentation and inverted sedimentation interesting dynamic states occur, which we shortly introduce now.

In the region colored in gray in fig.~\ref{fig:neutral_states}(a) we observe collections of sinking squirmers, which we call plumes. Although oriented upwards on average, they can sink due to the reduced viscous friction of a squirmer cluster. The plumes supply a convective roll at the bottom of the system that is formed and kept running by the self-pro\-pell\-ing squirmers. Solitary squirmer escape from the edges of the roll and swim upwards. 
The formation of rolls and plumes are reminiscent of bioconvection observed in ex\-peri\-ments~\cite{PedleyKessler1992,JanosiHorvath1998,HosoyaMogami2010,SatoToyoshima2018}. We show an example of this state at $\alpha=2.31$ and $r_0 / R\alpha =0.11$ in video M3 and also provide a snapshot in 
fig.~\ref{fig:neutral_states}(b), left. 
The density profile is non-monotonous with a broad maximum at the position of the bottom cluster, a minimum in the central region, where plumes pass through, and a sharp maximum at the top wall, where squirmers accumulate.

Notably, starting from an initially uniform distribution of squirmers we also observe plumes that form at the top wall, sink down, and then slowly evaporate. Likewise, convection rolls can be merely transient when the bottom cluster eventually disappears. The steady state for these cases are inverted sedimentation profiles with strong layering at the top wall. 
The transient plumes and rolls occur at higher torques in the blue region of the state diagram of fig.~\ref{fig:neutral_states}
beyond the dashed line and to the right of the  state of stable plumes and convection rolls.

An interesting situation arises in the state diagram when $\alpha$ is situated in a narrow stripe above $\alpha = 1$ for torques 
larger than a threshold value that we show by the lower solid line in fig.~\ref{fig:neutral_states}(a).
The clearest representation of this spawning-cluster state arises for large torques, where the squirmer orientation is fixed to the upright direction. A big cluster of squirmers floats above the lower wall [see fig.~\ref{fig:neutral_states}(b), right]. Hydrodynamic interactions between the squirmers increase their mobilities and thereby their sedimentation velocities, which can cancel the swimming  velocity even for $\alpha \ge 1$. To be more specific, a squirmer is pulled downward by the stokeslet flow fields from surrounding squirmers [\emph{cf.} eq.\ \eqref{eq:grav_flow_st}]. 
Taking 12 of them (hexagonal packing) at distances equal to the mean spacing in the spawning cluster gives a velocity roughly twice the swimming velocity $v_0$. Hence, squirmers in a cluster sink downward until they reach and interact with the bottom wall.
Figure\ \ref{fig:density_profiles} shows the sedimentation profile for $\alpha=1.5$ and $r_0 / R \alpha =0.5$.
In comparison to the convection roll the bottom cluster has a higher density visible by a more pronounced broad maximum and 
also the depletion in the middle of the cell is stronger by an order of magnitude. 
We call this state ``spawning cluster'' because individual squirmers occasionally escape from the pores within the cluster
at high velocity. This can be seen in video M4, as well as in the snapshot in fig.~\ref{fig:neutral_states}(b), right.

In table\ \ref{tab:states} we summarize the characteristics of the different states of neutral squirmers, including the figures which describe
them. In the following sections we discuss these states in more detail. We investigate conventional and inverted sedimentation of neutral squirmers in sect.~\ref{sec:sedimentation} and the state with plumes and convection rolls in sect.~\ref{sec:plumes}. Transient plumes and rolls in the inverted sedimentation state are discussed in sect.~\ref{sec:transients} and in sect.~\ref{sec:spawning} we 
address the spawning-cluster state. Finally, we show how the state diagram changes for pusher and puller squirmers in sect.~\ref{sec:flow_fields}. 

\subsection{Conventional and inverted sedimentation}
\label{sec:sedimentation}

\begin{figure}
\includegraphics[width=0.45\textwidth]{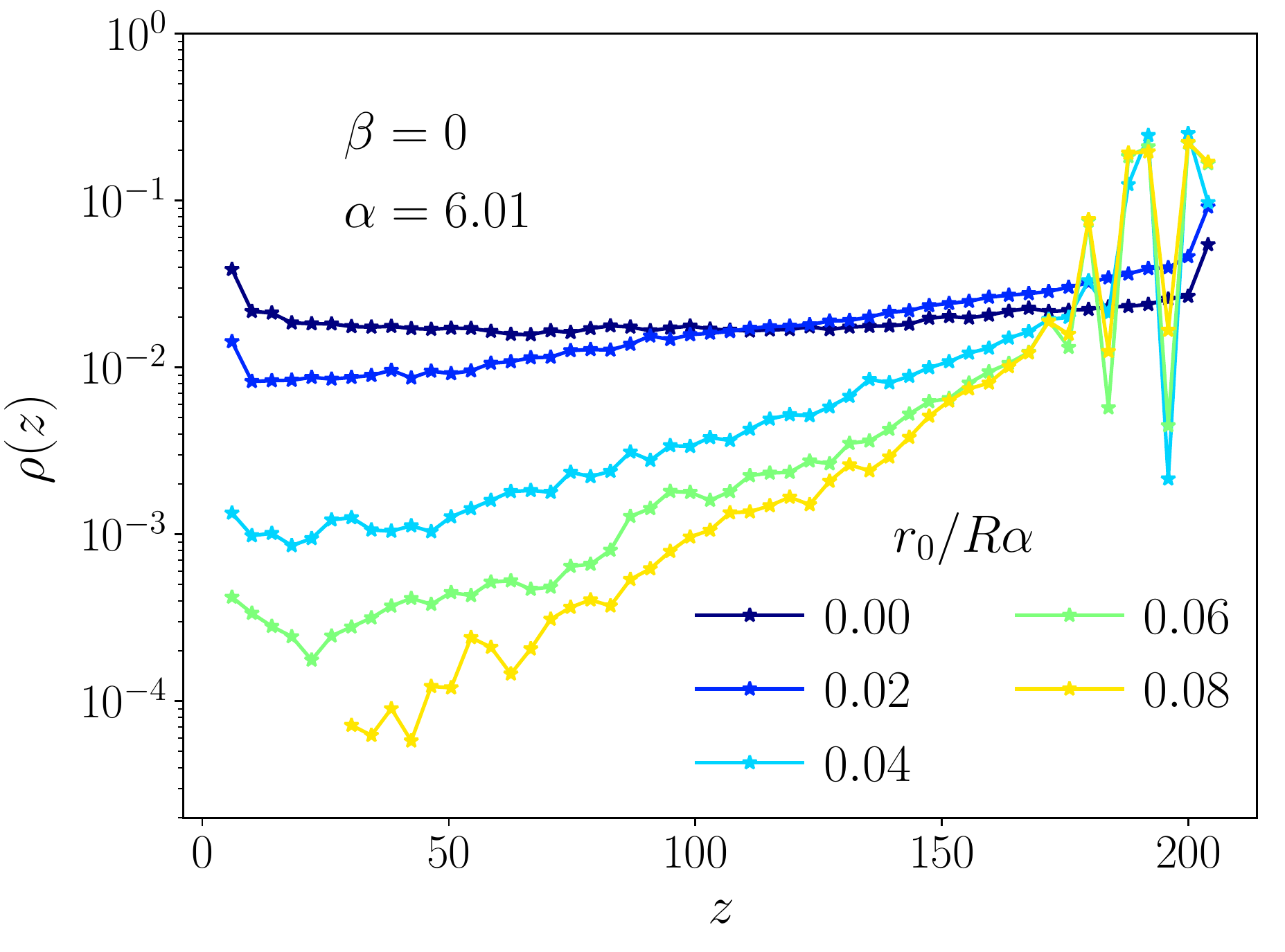}
\caption{Density profiles for the inverted sendimentation state for $\alpha=6.01$ and different torque values.
For high torques the steady state  is reached via a long-lived transient, where a cluster separates from the top layers and slowly evaporates (not shown).
}
\label{fig:g-005_density_profiles}
\end{figure}

\subsubsection{Sedimentation}
Collective sedimentation of squirmers under gravity has extensively been studied in refs.\ \cite{KuhrStark2017,KuhrStark2019,ShenLintuvuori2019}.
It also occurs, of course, for bottom-heavy squirmers. In the state diagram of fig.\ \ref{fig:neutral_states}(a) the sedimentation 
regime at low torques extends beyond $\alpha =1$, although single squirmers with upright orientation can overcome gravity.
For $\alpha=1.5$ and $r_0/R\alpha=0.01$ we already showed the exponential sedimentation profile in fig.~\ref{fig:density_profiles}.
It occurs because the flow fields of nearby squirmers tilt the orientation of one squirmer away from the upright direction, which  
therefore sinks even for $\alpha > 1$. The necessary flow vorticity $\boldsymbol{\omega} = \mathrm{curl} \, \mathbf{u} / 2$ for the reorientation is provided only by the grav\-i\-ty-in\-duced stokeslets 
[see eq.~\eqref{eq:grav_flow_st})]
since the flow field of neutral squirmers has zero vorticity.
The situation changes for large torques, where the squirmer orientation is always upright. At $\alpha < 1$ squirmers are confined to clusters sitting on the bottom wall. When crossing $\alpha \approx 1$ they start to float, in the sense that the density at $z=0$ develops a minimum. An exemplary density profile of these spawning clusters for $\alpha = 1.5$ and $r_0/R\alpha = 0.5$ was already discussed in connection with fig.\ \ref{fig:density_profiles}.

\subsubsection{Inverted sedimentation} 
In fig.~\ref{fig:g-005_density_profiles} we show a set of density profiles for the inverted sedimentation state for $\alpha=6.01$. While at zero torque the profile is nearly uniform, increasing the torque from zero the sedimentation length of the inverted profile
decreases and at high torques the inversion becomes strong enough that layers of squirmers form at the top wall. 
Note that for the three largest torque values transient plumes 
are observed. In concrete, 
a cluster separates from the top layers, sinks as a plume, and slowly evaporates (see video M8 in the supplemental
material, and sect.~\ref{sec:transients}). However, the sedimentation profiles are determined after reaching the steady state in the long-time limit.

\subsection{Plumes and convective rolls} 
\label{sec:plumes}
In the following we discuss squirmer plumes that constantly appear in the bulk and feed a convective roll at the bottom of the system. 

\begin{figure}
	\includegraphics[width=0.49\textwidth]{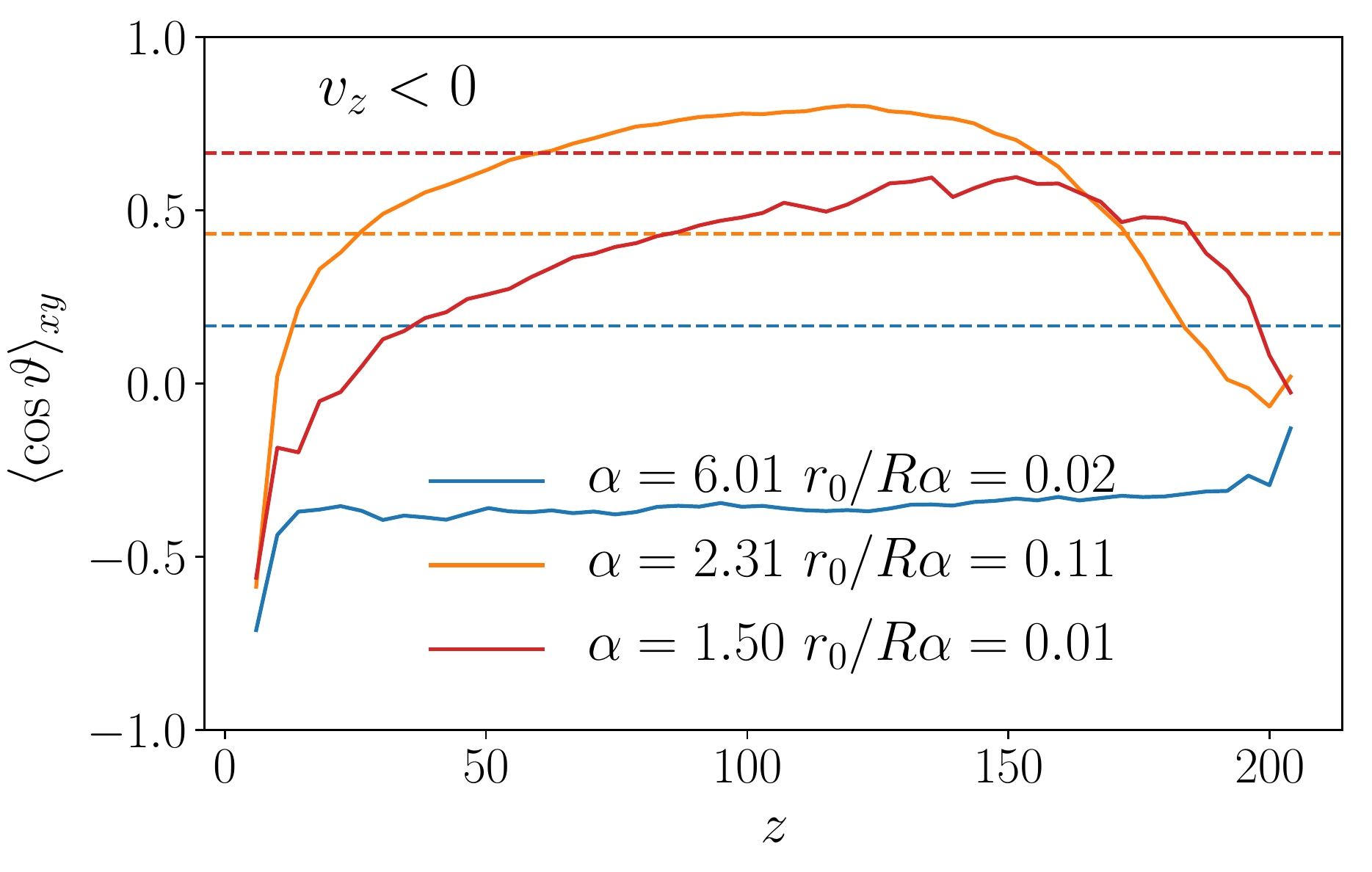}
	\caption{Mean vertical squirmer orientation $\langle\cos\vartheta\rangle_{xy}$ as a function of height $z$ for all squirmers with drift velocity  $v_z < 0$. The curves correspond to the states of sedimentation (red), inverted sedimentation (blue),
	and plumes and convective rolls (orange). Horizontal lines show $\cos\vartheta_{\text{th}}=1/\alpha$, where single squirmers 
	under gravity switch between up- and downwards swimming. 
}
	\label{fig:meanori}
\end{figure}

\subsubsection{Collective sinking and plumes}
\label{sec:collective_sinking}

Plumes of squirmers sink with preferentially upright orientation. To characterize this motional state and contrast it with
(inverted) sedimentation, we plot in fig.~\ref{fig:meanori} the mean vertical squirmer orientation $\langle\cos\vartheta\rangle_{xy}$ as a function of height $z$ for all squirmers, which drift downwards: $v_z<0$. The average is taken over the horizontal $xy$ plane.
In addition, for each $\alpha$ the dashed line indicates the threshold value $\cos\vartheta_\mathrm{th} = 1/\alpha$ for the degree of upright orientation, which a single squirmer must not exceed in order to sink.

In the conventional sedimentation state (red curve) the mean upright orientation is always below the threshold value, which
explains the downward drifting of the squirmers. The same is true for the inverted sedimentation state (blue curve).
However, while the threshold value $\cos\vartheta_\mathrm{th}$ belongs to a small upright orientation, the almost constant 
mean orientation is negative here, thus squirmers with velocity $v_z < 0$ swim downwards
(rather than sink).
In contrast, the mean orientation of squirmers in the plume state (orange curve) exceeds $\cos\vartheta_\mathrm{th}$ in the region away from the walls, where plumes occur. Thus, their sinking cannot be explained by looking at single squirmers. Instead, it occurs since hydrodynamic friction in clusters of squirmers is reduced such that the mobility of each squirmer is increased. Thus, their sedimentation velocity, which acts against the upwards swimming, is larger compared to single squirmers and the whole cluster can sink.
Indeed, for two squirmers the leading order flow field acting on the neighbor is given by the stokeslet contribution
in eq.\ \eqref{eq:grav_flow_st}, which provides a flow in negative $z$-direction and thus reinforces the gravitational sinking. This hydrodynamically induced mobility increase has already been studied for passive colloids on the basis of Rotne-Prager mobilities, for example, in refs.\ \cite{ReichertStark2004,ReichertStark2004II} and for larger conglomerates in refs.\ \cite{CichockiHinsen1995,LassoWeidman1986}.

\paragraph{Cluster velocities}

\begin{figure}
	\centering
		\includegraphics[width=0.49\textwidth]{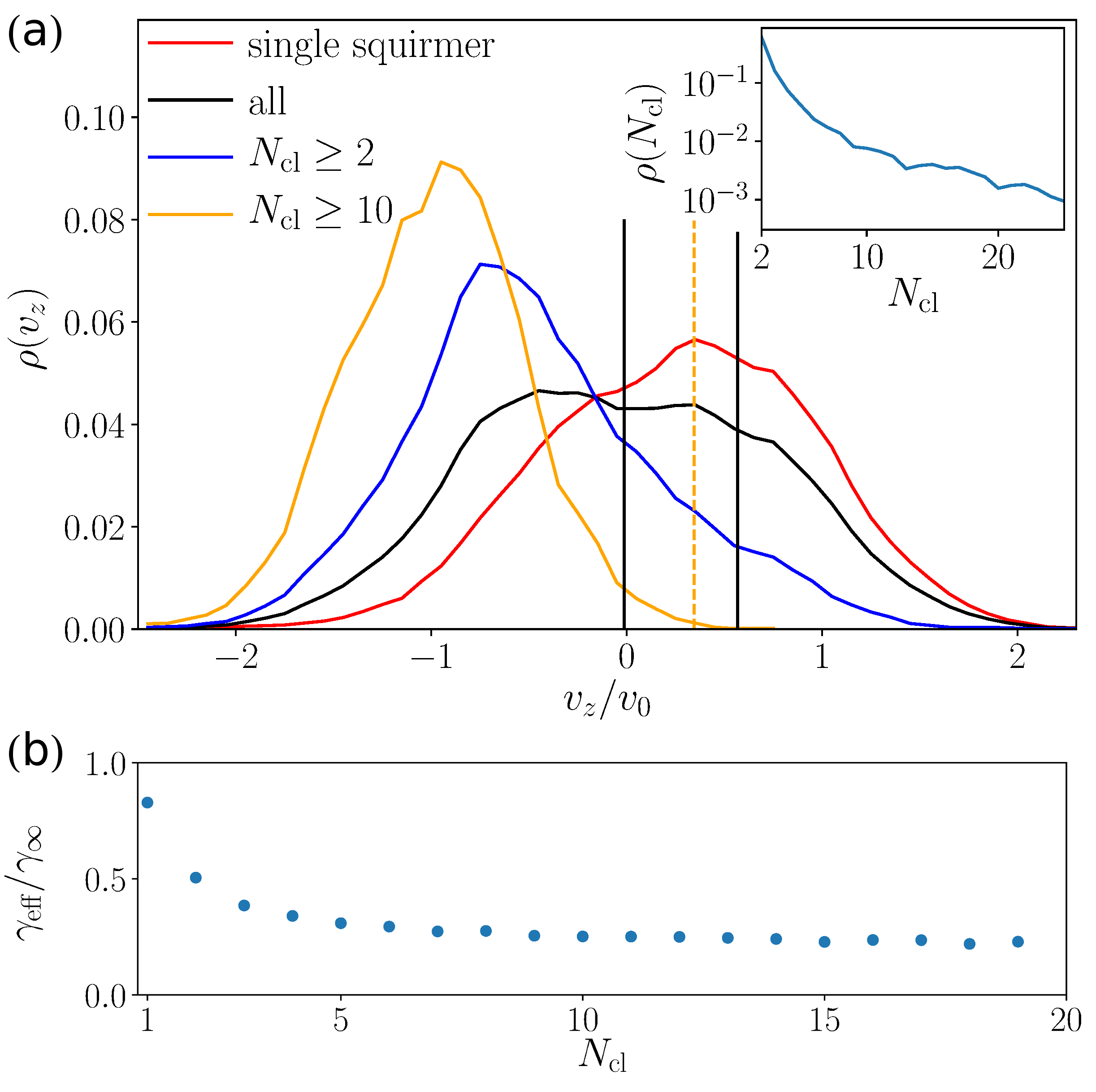}
		\caption{(a) Distribution of vertical velocities of neutral squirmers at $\alpha=2.31 $ and $r_0/R\alpha = 0.11$ for different cumulative cluster sizes $N_\mathrm{cl}$. Vertical lines: maximum bulk velocity $v=v_0-mg/\gamma_\infty$ (black, right), nearly zero mean velocity of all squirmers (black, left), velocity for the mean vertical orientation of all squirmers from $\rho(v_z)$ for $N_\mathrm{cl}\geq 10$ (orange). Inset: probability density of cluster sizes. (b) Effective friction coefficient experienced by squirmers in a cluster as a function of cluster size.}
		\label{fig:clusters}
\end{figure}

To quantify the collective sinking in the plume state further, we show in fig.\ \ref{fig:clusters}(a) the distributions
of the vertical cluster velocity for different cumulated cluster sizes $N_\mathrm{cl}$. To determine them, every 1000th
time step we monitor the clustering in the system by grouping squirmers that have a neighbor distance $d<R/8$ into the same 
cluster. This identifies clusters of different size $N_\mathrm{cl}$, where solitary squirmers have $N_\mathrm{cl}=1$. We only 
consider squirmers in a region with height from $60a_0 < z < 160 a_0$ in order to avoid the densely accumulated squirmers close 
to the top and bottom walls. The distribution of cluster sizes (excluding solitary swimmers) is shown in the inset of 
fig.~\ref{fig:clusters}(a).
Having grouped the squirmers into clusters of different sizes, we can determine the velocity distributions for all squirmers, solitary squirmers ($N_\mathrm{cl}=1$), for squirmers in clusters with $N_\mathrm{cl} \geq 2$, as well as those with $N_\mathrm{cl} \geq 10$ [see fig.\ \ref{fig:clusters}(a)]. As references we have  indicated three characteristic velocities by vertical lines: the maximum bulk 
velocity $v=v_0-mg/\gamma_\infty$ (right black line) and the zero mean velocity of all squirmers (left black line). 
The dashed orange line is the vertical velocity, which is calculated with the mean vertical orientation of all squirmers from the orange distribution with $N_\mathrm{cl}\geq 10$.

The total distribution of vertical squirmer velocities in the convective plume state has a broad shape symmetric about a mean very close to zero. This makes sense since in steady state there should not be a non-zero vertical flux of squirmers.
Solitary squirmers ($N_\mathrm{cl}=1$) have a broad distribution as well, but with a positive bias. 
This illustrates that solitary squirmers contribute more to upwards than to downwards motion.
Interestingly, the velocities of solitary squirmers and small clusters can exceed the free-swimming limit $v_z=v_0$. We speculate that this comes from squirmers approaching the convective roll at the bottom, where they are pushed up by hydrodynamic 
flow fields originating from the convective rolls.
Indeed, squirmers with $v_z > v_0$ are hardly present in the velocity distribution of an inverted sedimentation state (not shown)
where convective flows at the bottom do not exist.

Restricting ourselves to clusters of larger size, $N_\mathrm{cl} \ge k$, the distributions are more and more shifted to negative 
velocities with increasing $k$ and become narrower compared to both the single-particle and the overall distributions.
They also lose the tail with positive drift velocity $v_z$. The squirmer velocities in clusters are determined by the balance 
of self-propulsion and sedimentation velocities.
The latter is increased in clusters of squirmers compared to solitary squirmers due to their hydrodynamic interacions, as discussed above. 
In contrast, when one calculates for $N_{\text{cl}} \ge 10$ the mean vertical orientation and balances the upwards swimming and sedimentation velocities using the single-squirmer mobility, one obtains a small positive 
velocity indicated by the dashed orange line. It hardly intersects the distribution at its far end. However, this velocity is close to the mean of the single-squirmer velocity distribution (red curve), because we find that the orientational distributions of squirmers swimming alone or in clusters are very similar.
We attribute this to the missing vorticity in the flow field of a neutral squirmer.
Finally, for the system displayed in fig.\ \ref{fig:clusters}(a) at $N_\mathrm{cl}\ge 10$ the mean value of the vertical velocity is close to $-v_0$. It decreases further for even larger clusters and, of course, also depends on the velocity ratio $\alpha$.

In fig.\ \ref{fig:clusters}(b) the strong decrease of hydodynamic friction in clusters with increasing size is clearly visible. We estimated the effective friction coefficient $\gamma_\mathrm{eff} = -mg / v_{\text{eff}}$ by determining an effective sedimentation velocity $v_{\text{eff}}$ of a squirmer within a cluster. For this we averaged over
the sedimentation velocities $v_z - v_0\cos\vartheta$ of all squirmers within a cluster and subsequently took the mean over clusters with the same size that occur within a time window of $10^6 \Delta t$. 
We normalize the resulting effective friction by the bulk value. 
In addition to the strong size dependence
of $\gamma_\mathrm{eff}$, we realize how already for a single squirmer hydrodynamic interactions with its neighbors reduce the friction coefficient compared to $\gamma_\infty$ of an isolated squirmer.

\paragraph{Flow vorticity}
We have already mentioned how advection by hydrodynamic flow fields from neighboring squirmers (stokeslets in leading order)
enhances the sinking velocity and thus the mobility of squirmers in a plume. However, squirmers in the neighborhood of a sinking 
plume are also reoriented by the vorticity of these flow fields. While sinking, a squirmer reorients and therefore swims towards 
the plume. It joins the plume and thereby contributes to its  vertically extended shape. We note that such vorticities also play a role 
in the formation of a fluid pump by hydrodynamically interacting  active particles moving in a harmonic trap potential \cite{HennesStark2014}. 
In that study hydrodynamic torques only compete with rotational noise, whereas in the present case they have to balance the external 
gravitational torque acting on bottom-heavy squirmers. 

At high external torques stable plume states do not exist in the state diagram 
of fig.\ \ref{fig:neutral_states}(a), because bottom heaviness completely dominates the upright squirmer orientations. A reorientation 
by flow vorticity is not possible. However, when both hydrodynamic and gravitational torques are comparable, a neighboring 
squirmer tilts towards a plume but can only join it for a vertical alignment with $\cos\vartheta \lessapprox 1/\alpha$, because otherwise 
it will swim upwards as explained before. This is why plumes persist at higher torques for decreasing $\alpha$ in the state diagram of fig.~\ref{fig:neutral_states}(a). The mechanism reported here resembles an instability described previously \cite{PedleyKessler1992}, where reorientations by flow vorticity also induce the formation of plumes composed of gyrotactic algae.

To illustrate the balance of stokeslet vorticity and rotations induced by bottom heaviness, we consider two squirmers at the same height with a distance $r_{12}$. Taking the curl of eq.\ \eqref{eq:grav_flow_st} and setting it equal to
eq.\ \eqref{eq:angular_velocity_bh} leads to 
\begin{equation}
\frac{r_0}{R\alpha} \sin \vartheta = \frac{1}{\alpha} \left( \frac{R}{r_{12}} \right)^2
\label{eq.estimate}
\end{equation}
The second squirmer is tilted towards the first one, 
then moves towards it so that they can form a plume.
For a simple estimate in parameter space we choose the minimal distance $r_{12}=2R$ and the orientation $\theta=\pi/2$, where the bottom-heavy torque is maximal. The hyperbolic curve obtained from this estimate is shown in Fig.~\ref{fig:neutral_states}(a) 
and locates the region of the plume and convective roll state fairly well.

\paragraph{Sizes of sinking clusters}
\begin{figure}
	\centering
	\includegraphics[width=\linewidth]{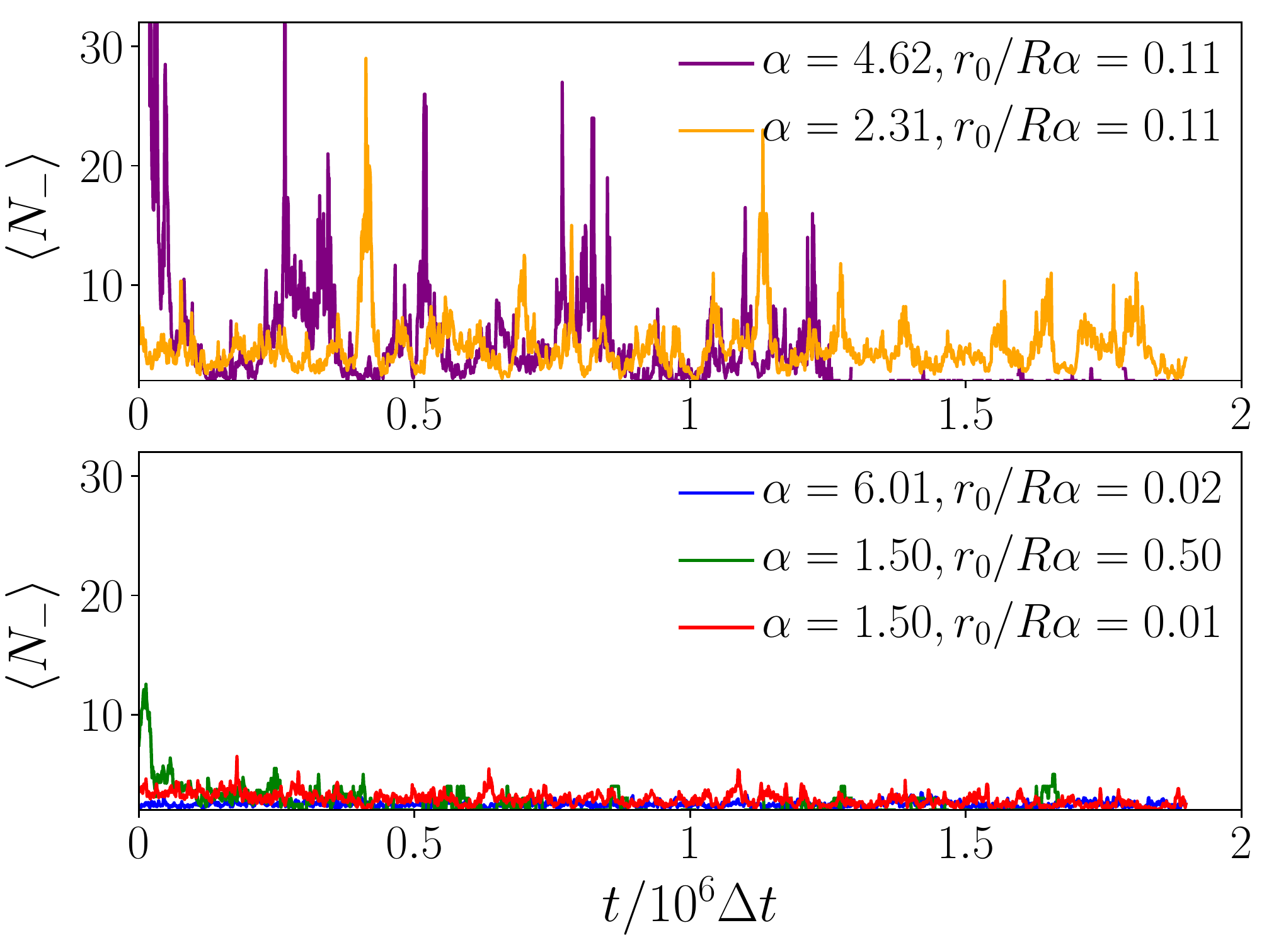}
	\caption{Mean number of squirmers in sinking clusters ($N_-\geq 2$) \textit{versus} simulation time in units of MPCD time step $\Delta t$. The plumes (orange) are clearly visible via distinct spikes. For transient plumes (purple) the spikes disappear with time. For conventional sedimentation (red), spawning clusters (green), and inverted sedimentation (blue) pronounced spikes are not visible.}
	\label{fig:mean_sinking}
\end{figure}

We have already identified the plumes in fig.~\ref{fig:meanori} via the mean vertical orientation of sinking squirmers. The clustering dynamics offers a further means to characterize and distinguish the stable plume state from the 
other states.
In fig.~\ref{fig:mean_sinking} we show the mean number of squirmers $\langle N_- \rangle$  in sinking clusters, \textit{i.e.}, where $N_- \geq 2$, over a period of $10^6$ MPCD time steps $\Delta t$ and within the same vertical region $60 < z/a_0 < 160$ as in fig.\ \ref{fig:clusters}. We can clearly recognize the plume state (orange line) by the high spikes that correspond to sudden events of collectively sinking squirmers passing through the region. 
Also shown is the transient plume state (purple line) that we discuss further in sect.~\ref{sec:transients}. Here, the spikes disappear around $1.3 \cdot 10^6 \Delta t$ when the plume has evaporated.
In contrast, in the inverted sedimentation state (blue line), the average size of sinking clusters remains low. Some small spikes 
can be seen for the conventional sedimentation state (red line), where small clusters form and induce some convective dynamics, which has been observed before~\cite{KuhrStark2017}. Likewise, some small spikes are visible for the spawning cluster state (green line) but much more rarely.

\subsubsection{Convective roll}
\label{sec:convective_roll}
\begin{figure*}[ht] 
	\centering
		\includegraphics[width=0.75\textwidth]{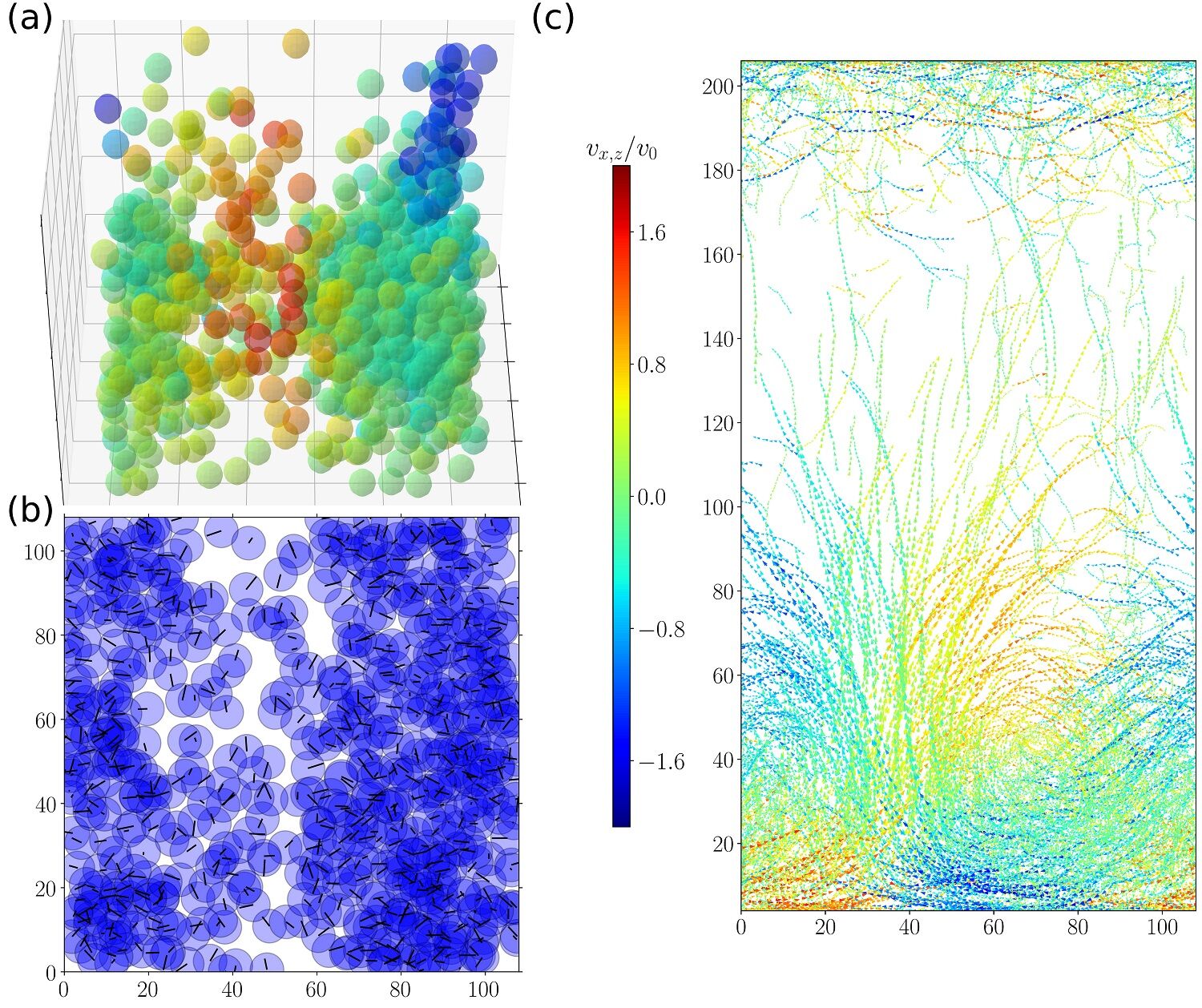}
	\caption{Convective roll in the bottom region of the system for the same parameters $\alpha=2.3$ and $r_0/R\alpha=0.11$  as in fig.~\ref{fig:neutral_states}(b). (a) 3D snapshot in the region $0\leq z/a_0 \leq 100$ with color-coded vertical 
          velocity components $v_z$. (b) Top view projection taking account of squirmers up to a height of $70a_0$.  (c) Cumulated
          squirmer positions showing squirmer trajectories projected on the vertical plane over a time span of roughly $9\cdot 10^5 \Delta t$ for the same viewing direction as in (a). The snapshots of (a) and (b) are included. The vertical velocity component $v_x$ is color-coded. The color bar applies to (a) and (c).}
	\label{fig:roll_snapshots}
\end{figure*}

\begin{figure}
	\centering
		\includegraphics[width=\linewidth]{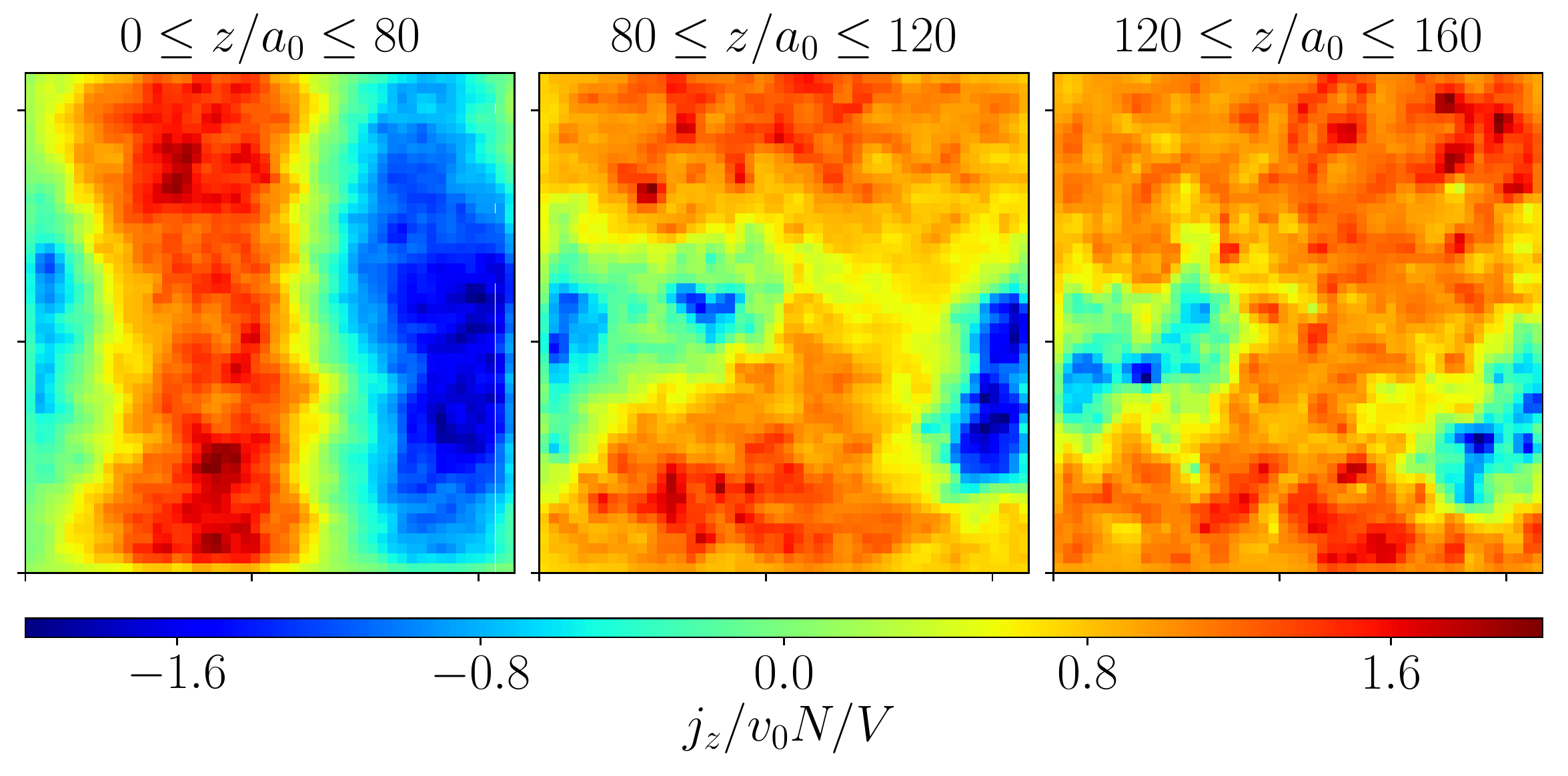}		
	\caption{Heat map of the mean vertical current density $j_z(x,y)$ in the horizontal plane and for three regions 
         $z/a_0\in [0,80]$, $[80,120]$, and $[120,160]$. The time range for averaging is the same as in fig.~\ref{fig:roll_snapshots}. We have applied a low-pass filter 
	(provided by python's scipy package) in order to smoothen the data.
}
	\label{fig:current_density}
\end{figure}

\begin{figure}
	\centering
	\includegraphics[width=0.95\linewidth]{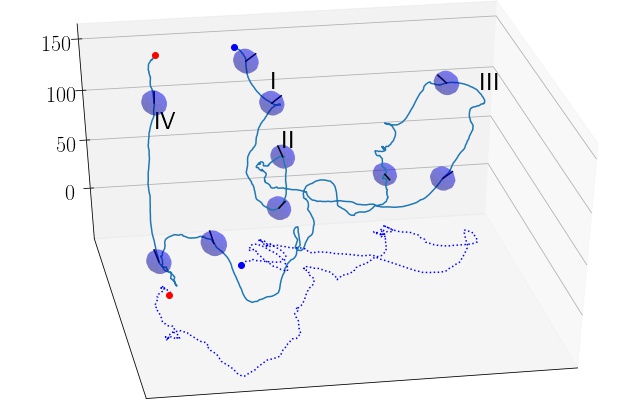}
	\caption{Three-dimensional 
    visualization of the trajectory of a spherical squirmer in the convective rolls. The squirmer first sinks as part of a plume (blue dot). Upon reaching the bottom cluster at position I it meanders laterally. Positions II and III indicate failed attempts to leave the cluster at the edge of a roll. Finally, it escapes at position IV and swims further upwards (red dot).
	Squirmers with their orientations are shown, their radii have been increased for better visibility.  Using the periodic boundary condition, the system box has been slightly shifted so that the squirmer does not leave the box. Dashed line: 
Projection of the trajectory on the $x$-$y$ plane.}
	\label{fig:highlight_3d_traj}
\end{figure}

In the context of microswimmers, convection rolls are stationary rotational patterns, which are formed by gravitactic swimmers 
due to their self-propulsion and advection in the self-generated flow fields \cite{JanosiHorvath1998,CzirokKessler2000,KruegerMaass2016}. 
Here, we use this term for the recirculating motion inside a cluster sitting at the bottom wall.
The cluster is visible in the left snapshot of fig.~\ref{fig:neutral_states}(b) and  in video M3. In 
video M5, we show a 3D view of the system in the region $0 \leq z/a_0 \leq 100$, where we have color-coded the vertical squirmer
velocities. A snapshot from the video is depicted in fig.~\ref{fig:roll_snapshots}(a). 
A very dynamic situation is visible. Squirmers in a plume sink down on the right and join the cluster. Because of their negative 
velocities they are colored in blue [see color bar in fig.\ \ref{fig:roll_snapshots}(c)]. The plume is also visible on the left due to 
the periodic boundary condition. At the same time, individual squirmers colored in red swim upwards in a region with low density. In fig.~\ref{fig:roll_snapshots}(b) we see this depleted region more clearly from the top. All squirmers below $z=70a_0$ are plotted.
Furthermore, we recognize that the convective roll has an elongated shape. This will change for larger systems as we demonstrate below. 

The circulation pattern in the convective roll is visualized in fig.~\ref{fig:roll_snapshots}(c). Here, we plot all the trajectories of squirmers within a slice of thickness $\Delta y = R$ spanning roughly the time interval $9\cdot 10^5 \Delta t$.
Each velocity vector of a squirmer is colored according to the value of its horizontal velocity component $v_x$.
We recognize two wavy patterns in the bottom region below $z/a_0 = 100$. They originate from squirmers moving 
down- and upwards while swimming to the left (blue lines) and from squirmers moving up- and downwards while swimming 
to the right (red lines). This generates the two circular patterns of the rolls. 
One is clearly visible in the right region, a second one on the left is not complete due to the periodic boundary condition. 
In the middle we see squirmers leaving the convective rolls.

In order to study the vertical motion in the system, we plot in fig.\ \ref{fig:current_density} the mean vertical current density of the squirmers $j_z(x,y)$, defined via $j_z(x,y):= \overline{\langle\rho(\mathbf{r})\rangle_\parallel \langle v_z(\mathbf{r})\rangle_\parallel}$
\cite{KuhrStark2017}, where $\rho$ is the squirmer density, $\langle \dots \rangle_\parallel$ means
average along the $z$-direction, and $ \overline{ \cdots}$ average over time. Figure\ \ref{fig:current_density} shows the current density with the vertical average taken in three different regions: at the botton where the convective rolls are ($0 \le z/a_0 \le 80$), in the middle where the plumes predominately occur ($80 \le z/a_0 \le 120$), and in the  bulk region above it ($120 \le z/a_0 \le 160$).
At the bottom (left plot) we immediately recognize the low density region where single squirmers move up (red region), while the two counterrotating rolls meet where the squirmers drift downwards (blue region). In the middle section (center plot) the sinking squirmers are spatially focussed due to the plume formation:  squirmers swim towards each other and sink collectively. 
The plumes feed the two counterrotating rolls and, therefore, are found at the right edge.
In contrast, rising squirmers move mainly individually, which is why their distribution is more spread out (center and right plot). In the middle height section (center plot) a weak plume is observable close to the center of the plane. In the videos M3 and M6 we see how such plumes move to the side while sinking. This exemplifies the strong hydrodynamic flows that uphold the convective roll.

While the two convective rolls convey the picture of a regular structure, the path of a single squirmer inside the dense cluster
is irregular, as we show in fig.~\ref{fig:highlight_3d_traj}. We have reoriented the point of view in comparison to fig.~\ref{fig:roll_snapshots}(a) by $90^{\circ}$, 
such that we now look at the long side of the rolls. Furthermore, taking into account the periodic boundary condition,
we have slightly shifted the system box so that the squirmer does not leave the box. For special points in time we also display the squirmer's orientation vector in the figure. The trajectory starts at the blue dot. After joining the rolls 
at position I, the squirmer meanders inside  the dense cluster and eventually escapes from an edge (IV) with an upright orientation. The plotted trajectory ends at the red dot.
The squirmer attempts to leave the bottom cluster several times (II,III) but is unsuccesful because the orientation is tilted too strongly against the vertical. It even crosses the low density region close to position IV as the projected trajectory on the horizontal plane shows. The strong variation of the squirmer orientation against the normal is, of course, due to the vorticity of the generated hydrdynamic flow field. In our case, it results dominantly from the stokeslet due to the gravitational force acting on each squirmer: The source-dipole far-field of the neutral squirmer has zero vorticity, while wall-induced image fields decay with $\mathcal{O}(r^{-4})$ and therefore are weak \cite{RuehleStark2018,SpagnolieLauga2012}.
Interestingly, the  ``minuet'' dance of a pair of Volvox algae in the experiments of ref. \cite{DrescherGoldstein2009} could also be reproduced by stokeslet interactions between the two microswimmers. The more chaotic meandering motion observed here inside the convective rolls then occurs due to the interactions with all the surrounding squirmers.

\begin{figure}
	\includegraphics[width=0.48\linewidth]{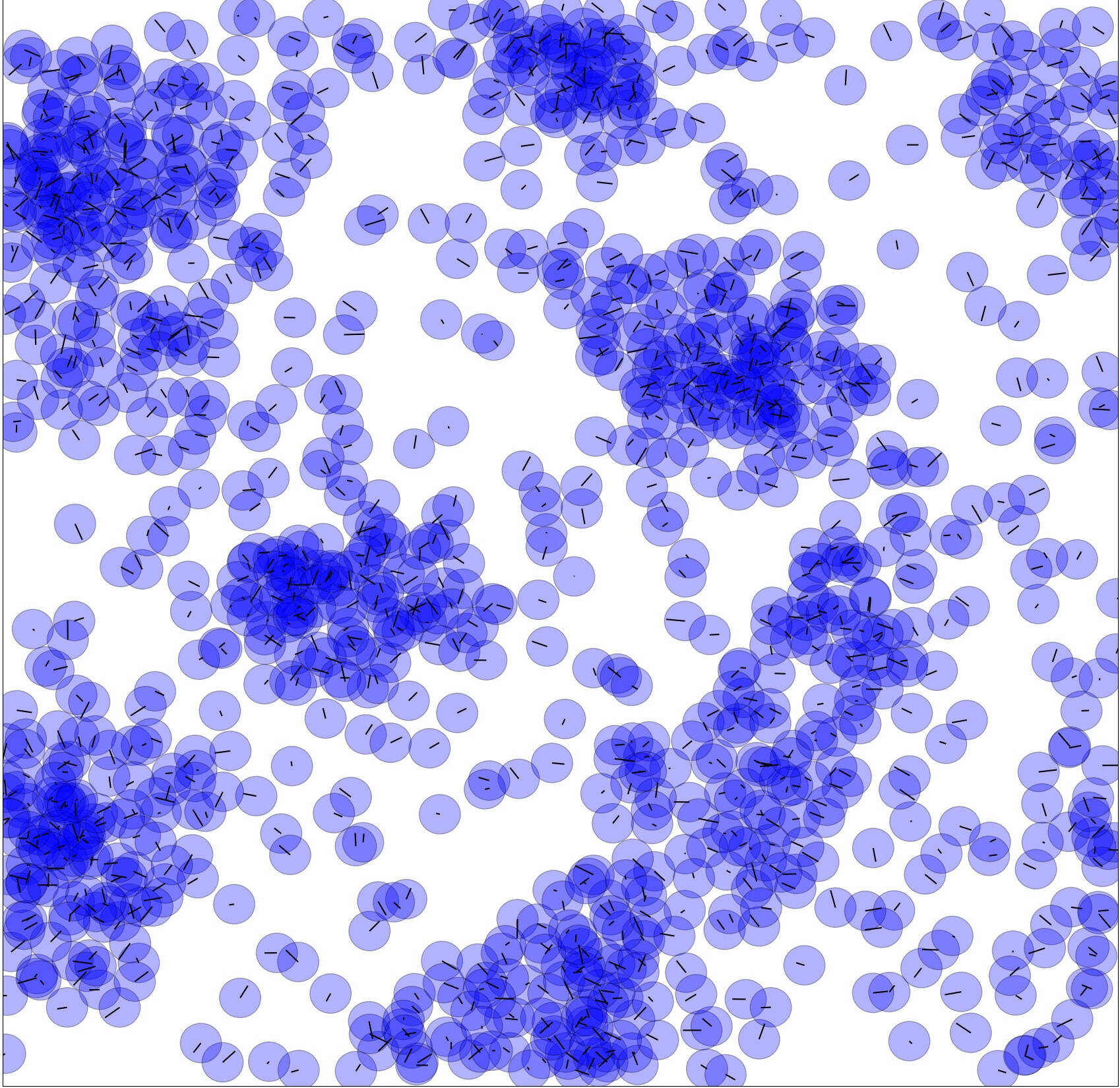}\hfill
	\includegraphics[width=0.48\linewidth]{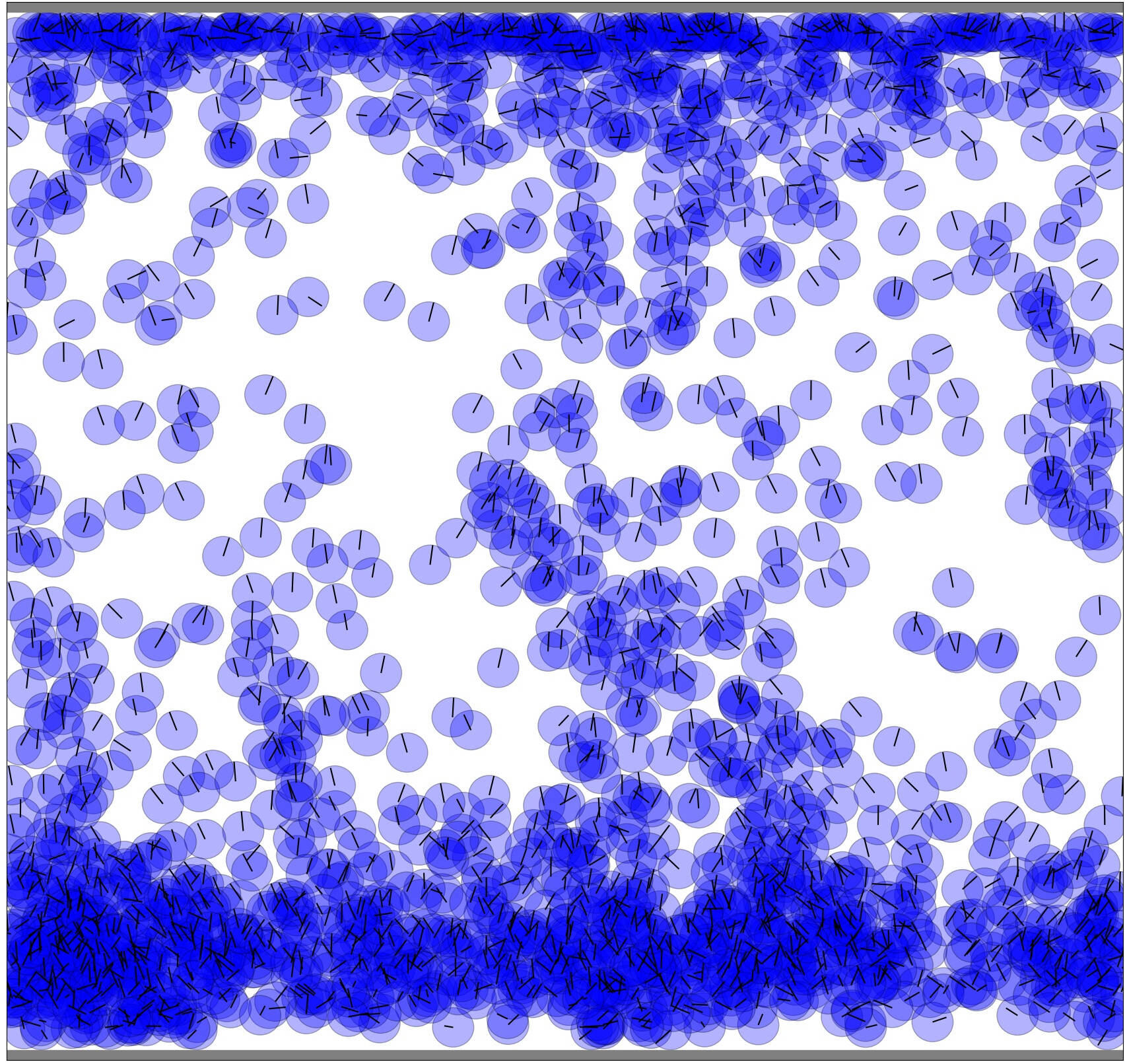}
	\caption{Snapshots for an array of convective rolls in a system with a horizontal cross-section that has been increased by a factor of four in comparison to fig.~\ref{fig:roll_snapshots}, while keeping the height constant. Left: Top view showing all squirmers below $z=100a_0$. Right: Side view.}
	\label{fig:larger_system}
\end{figure}

We already mentioned that the elongated convective rolls occupy the whole horizonzal plane of our simulation box. To avoid this finite-size effect, we double the edge length of the square cross section, keep the height constant, but reduce the squirmer volume fraction by a factor of two to 5\%. 
As the top view of fig.~\ref{fig:larger_system}, left demonstrates, several compact convective rolls with an island shape appear
in contrast to a single elongated cluster. In the side view of fig.~\ref{fig:larger_system}, right we re\-cog\-nize several sinking plumes. Videos M6 and M7 illustrates impressively the observed dynamics, in particular, how sinking plumes feed the convective 
rolls. Interestingly, rolls in the larger system have a characteristic distance. Increasing the density to 10\% as in the small systems 
the islands become larger and partially touch each other, but the distance of their centers stays approximately the same. Such 
regular structures are known in biological systems, where length scales can vary with system height or density, depending on the  
swimming mechanisms and vertical alignment~\cite{BeesHill1997,CzirokKessler2000}.
However, a systematic variation of system size or density is beyond the scope of this article.

\begin{figure}[t!]
\centering
	\includegraphics[width=0.85\linewidth]{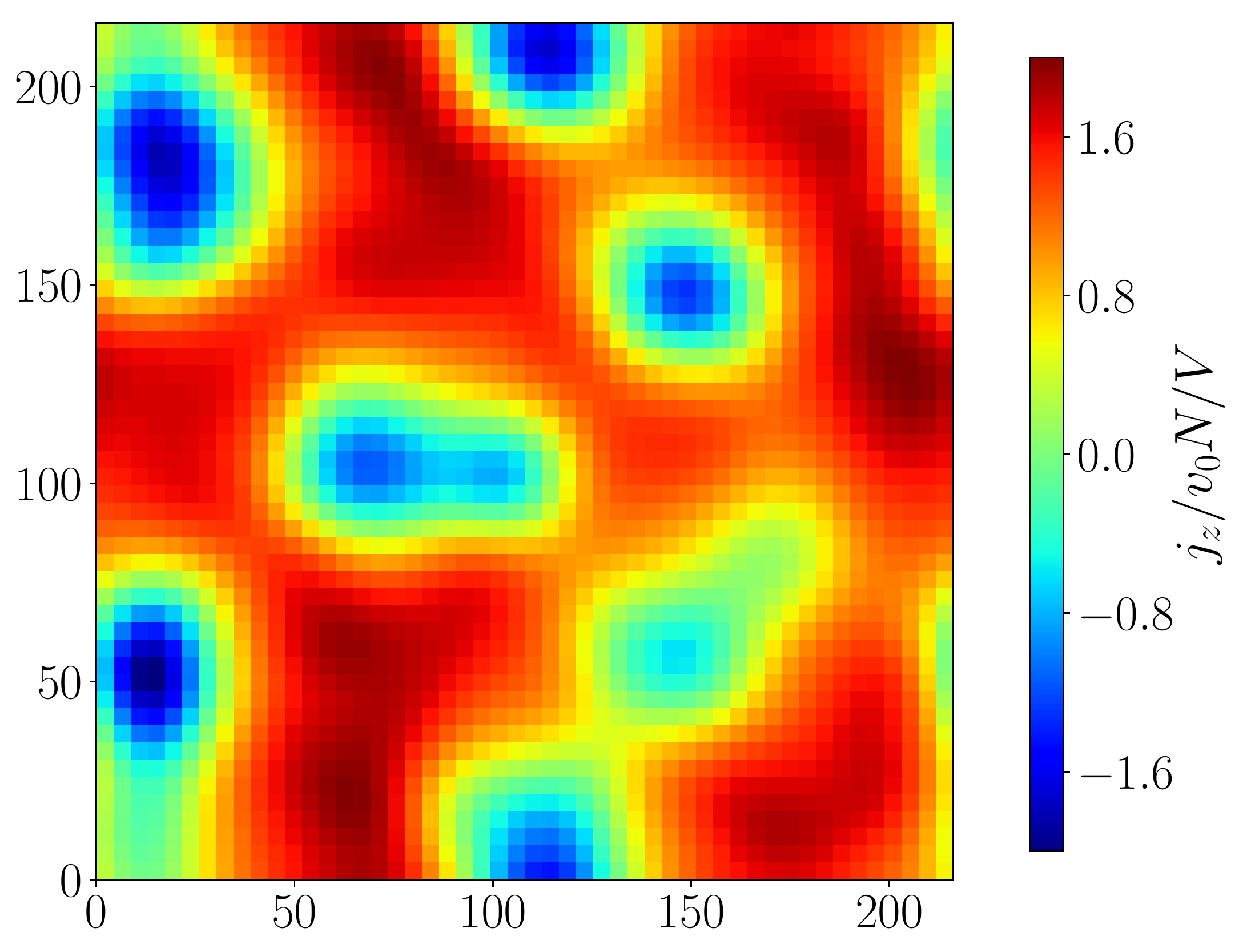}
	\caption{Mean vertical current density for the same system size as in fig.\ \ref{fig:larger_system}.
         The vertical region for determining the average is 
         $z/a_0\in [0,100]$. 
         The chosen time span of $4.5\cdot 10^5\Delta t$  includes the snapshots  of fig.~\ref{fig:larger_system}.  
         A low-pass filter (provided by python's scipy package) was applied in order to smoothen the data.}
	\label{fig:larger_current}
\end{figure}

\begin{figure}
\centering
	\includegraphics[width=0.49\linewidth]{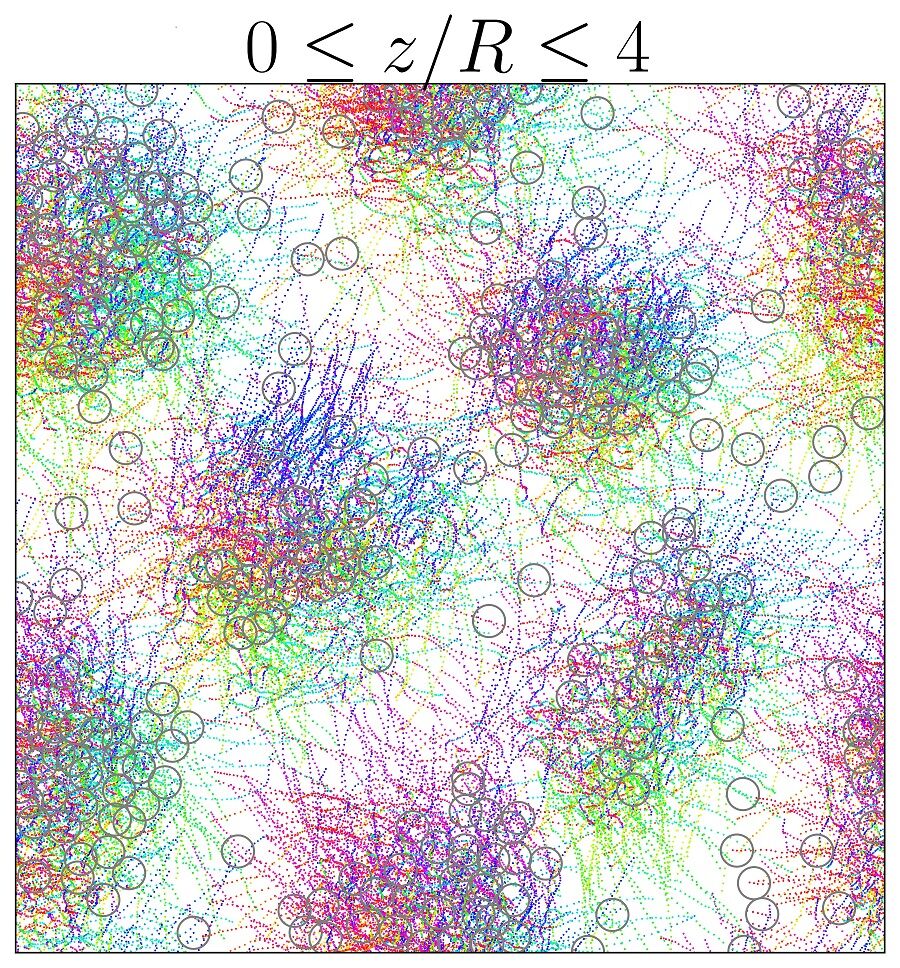}
	\includegraphics[width=0.49\linewidth]{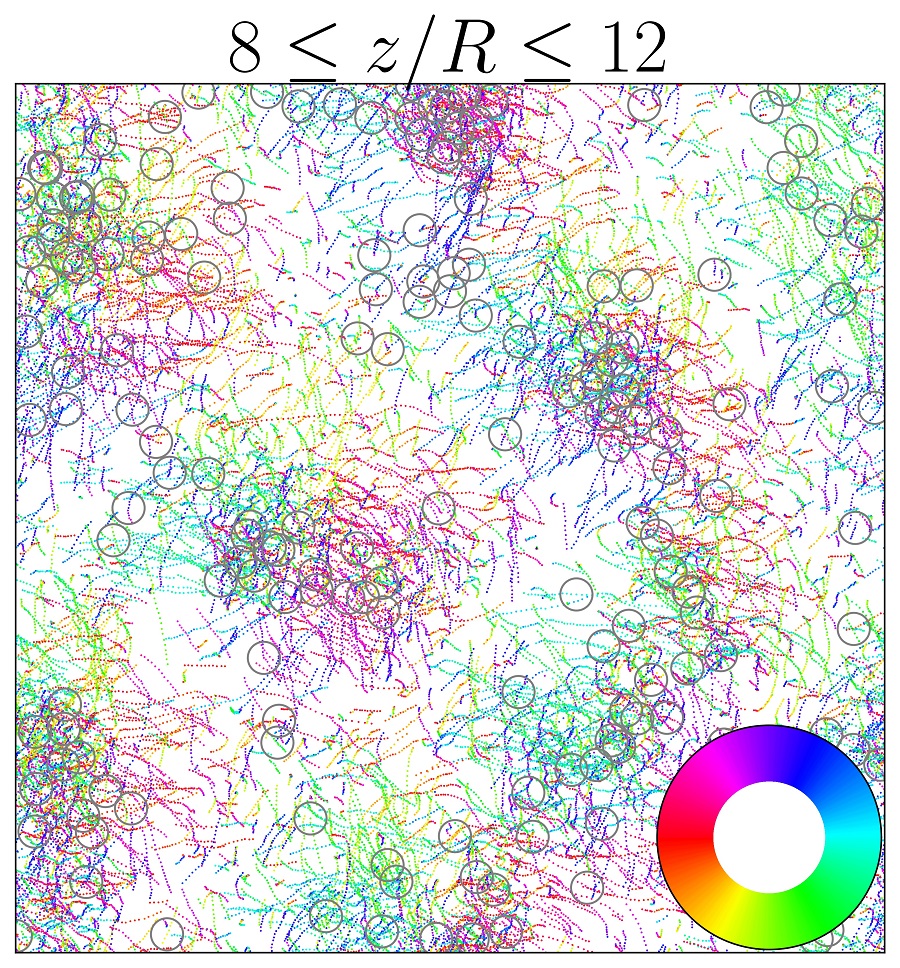}
	\caption{
	Squirmer trajectories during a time span of  $4\cdot 10^5 \Delta t$ projected onto the horizontal plane.
	Left: for the bottom vertical region of the convective rolls at $0 \le z/R \le 4$. Right:  for the upper vertical region at 
	$8 \le z/R \le 12$. The horizontal velocity direction is color-coded according to the circular color bar. The 
	faint circles show squirmers from the snapshot in fig.~\ref{fig:larger_system}, left within the respective vertical regions.
    \label{fig:larger_vhorizontal}        
}
\end{figure}

We clearly see the island shape of the convective rolls also in the mean vertical current density illustrated in 
fig.\ \ref{fig:larger_current}. The sinking plumes above the rolls and the sinking squirmers inside them are visible by the
blue areas, while around these regions squirmers swim upwards.
Figure\ \ref{fig:larger_vhorizontal} completes the picture of the convective roll. We show squirmer trajectories inside horizontal slices of thickness $4R$ either at the bottom (left) or top (right) of the convective rolls. The
direction of the in-plane velocity is color-coded. At the bottom squirmers move radially outwards (away from the dense clusters)
while at the top they move radially inwards (towards the centers of the dense clusters).
Combined with the vertical squirmer current, this implies a toroidal flow pattern for the convective rolls, which is also visible in video M7.

\subsection{Transient plumes and rolls}
\label{sec:transients}

Plumes and convective rolls as described in sect.~\ref{sec:plumes} are not stable at high torques. Instead, the long-term steady state of the system is an inverted exponential sedimentation profile that develops after a long-lived transient. In the following, we distinguish between two different transient states, which we observed starting from an initially uniform distribution of squirmers: evaporating plumes forming at the top wall and unstable rolls.

\subsubsection{Evaporating plumes}
\label{sec:evaporating}
First, we consider systems with large $\alpha\gtrapprox 5.6$. In the state diagram of fig.~\ref{fig:neutral_states}(a) we have identified inverted sedimentation with transient plumes to the right of the dashed line. At such high velocity ratios $\alpha$ stable plumes cannot exist since the vorticity from the gravitational stokeslet is too weak to orient neighboring squirmers towards each other in order to form stable plumes while sinking.
Instead, squirmers escape to the top wall already for external torques of medium strength. Hence, starting from a uniform distribution dense squirmer layers start to form at the top wall.
Here, due to flow vorticity the orientations tilt and protrusions of squirmers form that eventually separate from the layers. The sinking squirmer cluster reaches a final height well above the bottom wall. It emits squirmers which join the layers at the top wall. Thus the plumes gradually evaporate and disappear. The whole process can be seen in video M8. In steady state inverted sedimentation profiles with a few layers occur. We have already plotted some of them in fig.\ \ref{fig:g-005_density_profiles} for $r_0 / R\alpha \ge 0.04$ and also show such a profile at the end of video M8. The inability of a sinking cluster to attract more squirmers to support itself shows how gyrotactic structure formation fails for squirmers at high $\alpha$. As a consequence the steady  state here is always an inverted sedimentation for any torque.
Note, gravitational detachment of protrusions from a top layer has been extensively studied in different theoretical and experimental settings in connection with bioconvection\ \cite{PlessetWinet1974,ChildressSpiegel1975,PedleyKessler1992,JanosiHorvath1998,SatoToyoshima2018}.

\begin{figure}
	\includegraphics[width=\linewidth]{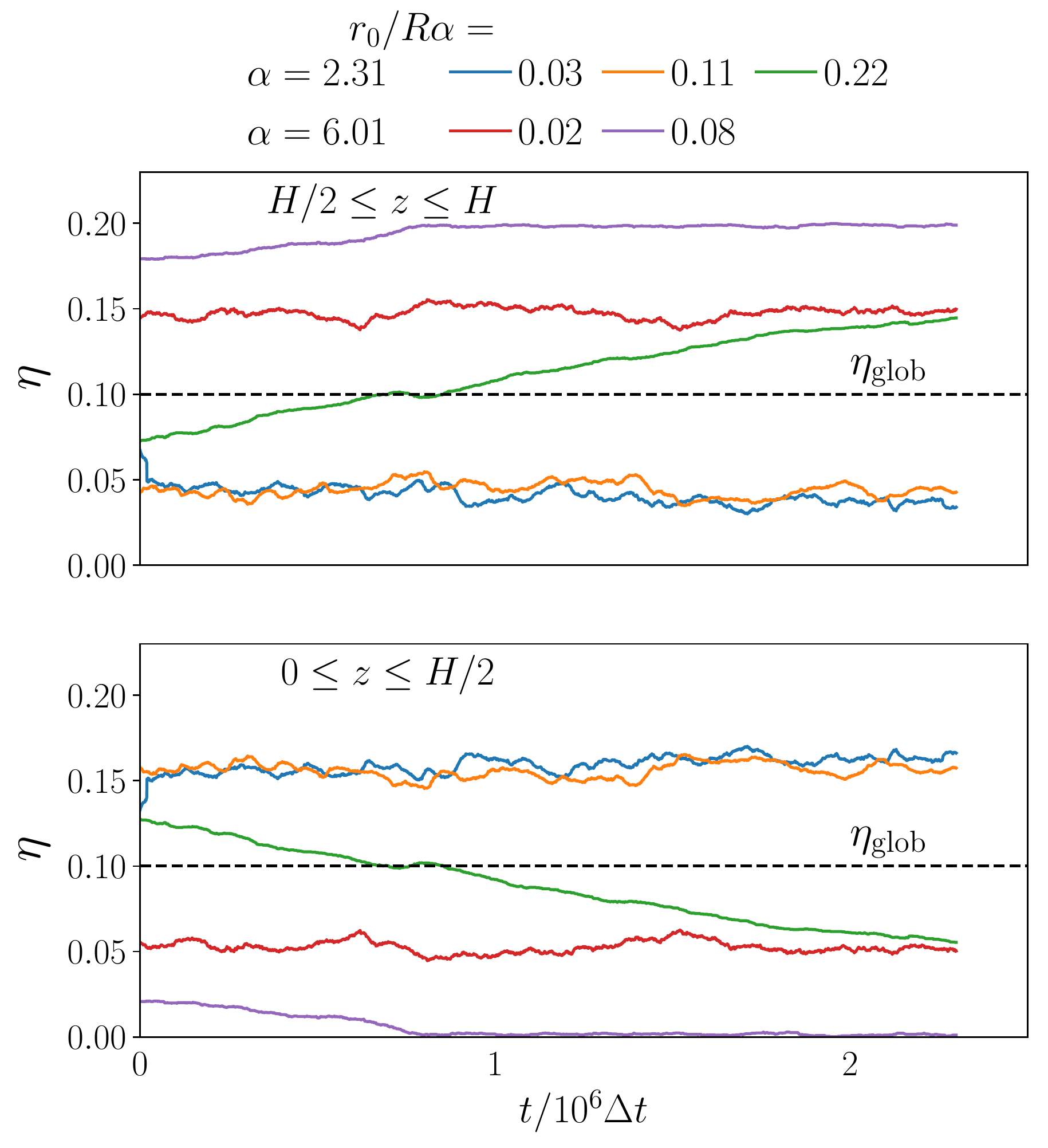}
	\caption{Time evolution of the volume fraction $\eta$ in the lower ($0 < z < H/2$) and upper  ($H/2 < z < H$) half of the system. The height is $H=210a_0$. The dashed line denotes the global volume fraction $\eta_\mathrm{glob}=0.10$. The curves correspond to inverted (red) and conventional sedimentation (blue), stable plumes and convective rolls (orange), transient rolls (green), and transient plumes (purple).
}
	\label{fig:volume_fraction}
\end{figure}

\subsubsection{Transient convective rolls}
We now consider lower values of $\alpha$,  where sinking plumes and convective rolls form, starting from the uniform initial distribution. Now, an increasing external torque dominates the orientational dynamics of squirmers meaning they experience a stronger vertical bias. This counteracts the hydrodynamic reorientation and the motion of squirmers towards a plume. Thus, at higher torques the plumes become thinner and eventually disappear in the long-time limit. 
As a result, the convective rolls at the bottom wall do not receive sufficient influx of squirmers and also evaporate.
Again, the system reaches a steady state with an inverted sedimentation profile and with layering at the top wall.
The process is visualized in video M9.

Both of the observed transient structures reveal themselves with clear signatures in the time evolution of the
spatial squirmer distribution. In fig.\ \ref{fig:volume_fraction} we plot the volume fraction $\eta$ versus time, for both the lower and upper half of the system ($z \in [0,H/2]$ and $[H/2,H]$, where $H$ is the system height). 
The curves for the steady states of inverted and conventional sedimentation (red and blue curves), as well as for stable plumes and convective rolls (orange curve) fluctuate around a constant value.
However, transient plumes sinking from the top wall (purple curve) or transient convective rolls (green curve) have a steadily decreasing density in the lower half of the system. At the same time, the squirmer layers at the top wall grow at the expense of the shrinking plumes and rolls. Simulating the transients is very time-consuming. For example, the green curve belongs to a transient convective roll and has not equilibrated yet. Initially, the density is high in the lower half of the system and the roll takes a long time to dissolve.

\subsection{Spawning clusters and transient hovering} 
\label{sec:spawning}

\begin{figure}
	\centering
	\includegraphics[width=0.45\linewidth]{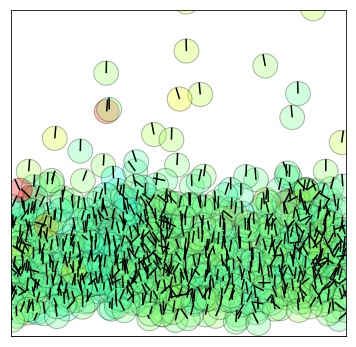}
	\includegraphics[width=0.45\linewidth]{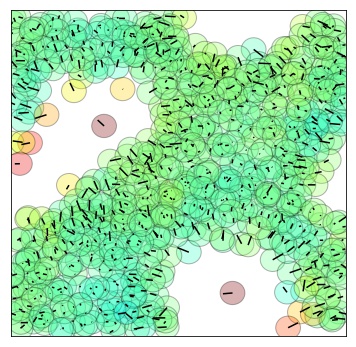}
	\caption{Left: Side-view snapshot of a spawning cluster at $\alpha=1.5, r_0/R\alpha=0.5$
	The cluster has a height of ca. $50a_0 = 12.5R$. Right: Top-view snapshot of the same system. Vertical velocities are color-coded with the same scale as in fig.~\ref{fig:roll_snapshots}.}
	\label{fig:spawning_snapshots}	
\end{figure}
\begin{figure}
	\centering
\includegraphics[width=\linewidth]{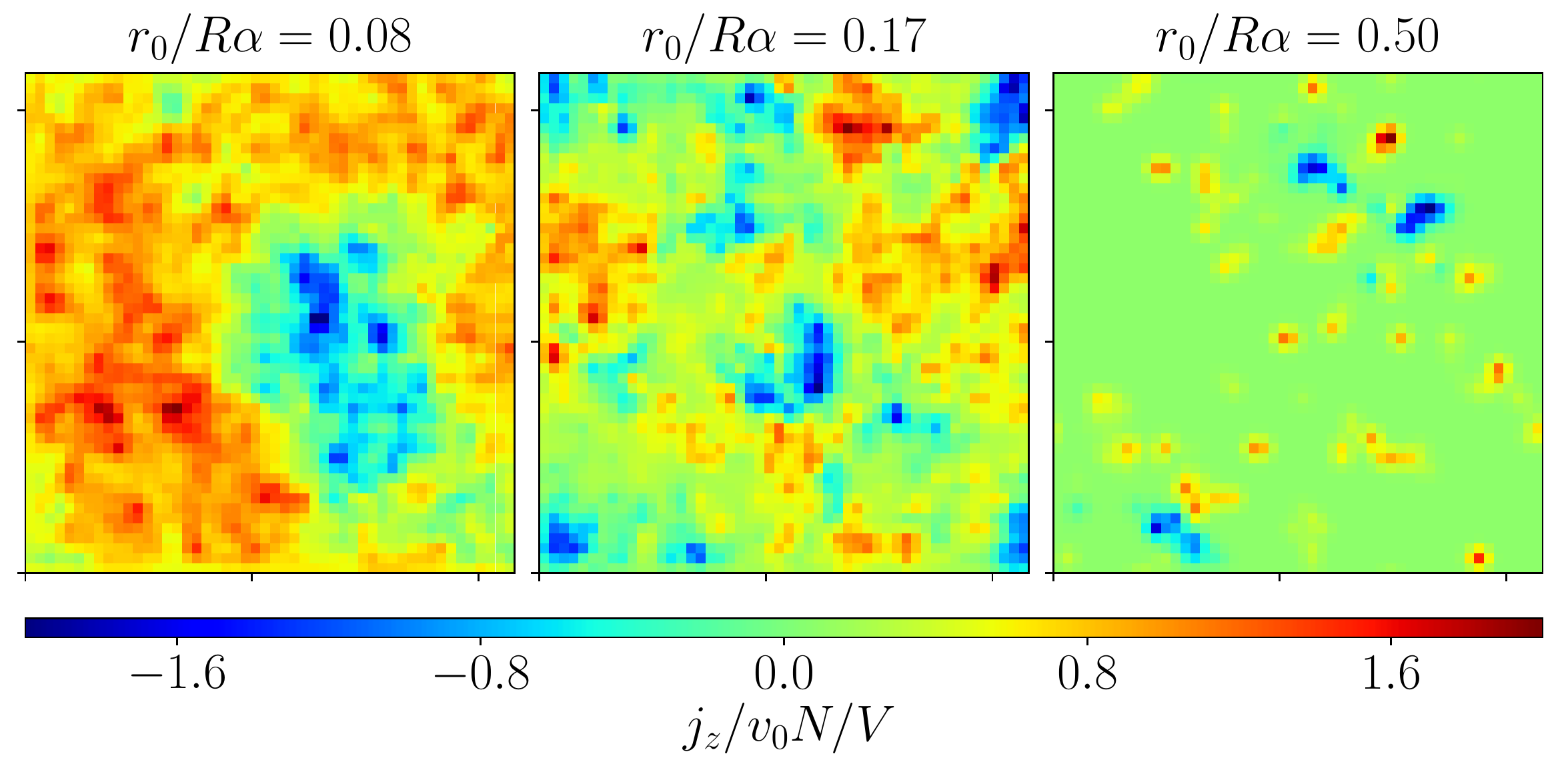}
	\caption{Heat map of the mean vertical current density $j_z(x,y)$ in the horizontal plane at $\alpha = 1.5$ in the region $80\leq z/a_0 \leq 120$ and averaged over a time period of $9\cdot 10^5\Delta t$. 
The torque value $r_0/R\alpha$ corresponds to sedimentation (left) and a spawning cluster (middle  and right). We have applied a low-pass filter (provided by python's scipy package) in order to smoothen the data.}
	\label{fig:spawning_2dhist}
\end{figure}

In fig.~\ref{fig:spawning_snapshots} we show a spawning cluster viewed from the side (left) and from the top (right). 
The snapshots belong to a spawning cluster at $\alpha=1.5$ and large torque situated at the far right of the state diagram in fig.~\ref{fig:neutral_states}.
We also show this system in video M4.
In the top view of the right snapshot we observe a porous structure of the cluster. In contrast to the convective rolls discussed above, holes strongly depleted by squirmers are visible. Thus the clusters are not compact objects. 
We colored the squirmers in the snapshots according to their vertical velocity $v_z$ using the same 
color code as in fig.~\ref{fig:roll_snapshots}. 
Hence, the green color of most squirmers shows that they move little in the vertical direction. Single squirmers perform a random walk or meander around within the cluster and when they reach the edge of a hole, the flow field of
the neighboring squirmers strongly drifts them upwards with large velocities up to $3 v_0$. They either rejoin the cluster or leave it. This is nicely visible in video M4.

In the spawning-cluster state plumes and convective rolls no longer exist, as videos M4 and M7 demonstrate and
when inspecting spawning clusters at different parameters. This becomes also clear from fig.~\ref{fig:spawning_2dhist}, where we show the mean vertical current densities in the region $z/a_0 \in [80,120]$ above the spwaning clusters and for three different torques at $\alpha =1.5$.

The left plot for $r_0/R\alpha = 0.08$ represents the sedimentation state. Similar to ref.~\cite{KuhrStark2017} convective patterns
in the region with an exponential sedimentation profile are visible, which consist of clearly separated areas with upward and downwards moving squirmers. Increasing the rescaled torque to $0.17$ (see video M10), these areas start to disintegrate
and thereby mark the onset of the spawning-cluster state. Increasing the torque even further to $0.50$, the mean vertical current density is zero everywhere except for some small patches. This is 
consistent with the very small density in the bulk region of our system as demonstrated by the corresponding density profile in fig.~\ref{fig:density_profiles}. Note, however, that compared to the sedimentation state the spawning-cluster state has a much higher density at the top wall, where squirmers leaving the cluster gather.

In the state diagram of fig.\ \ref{fig:neutral_states} we also mention ``transient hovering''. For very large external torques, well above the values used in the state diagram, we expect spawning clusters to dissolve with time at $\alpha > 1$. However, close to $\alpha = 1$ they dissolve very slowly.

\begin{figure*}[t!!]
\includegraphics[width=0.49\textwidth]{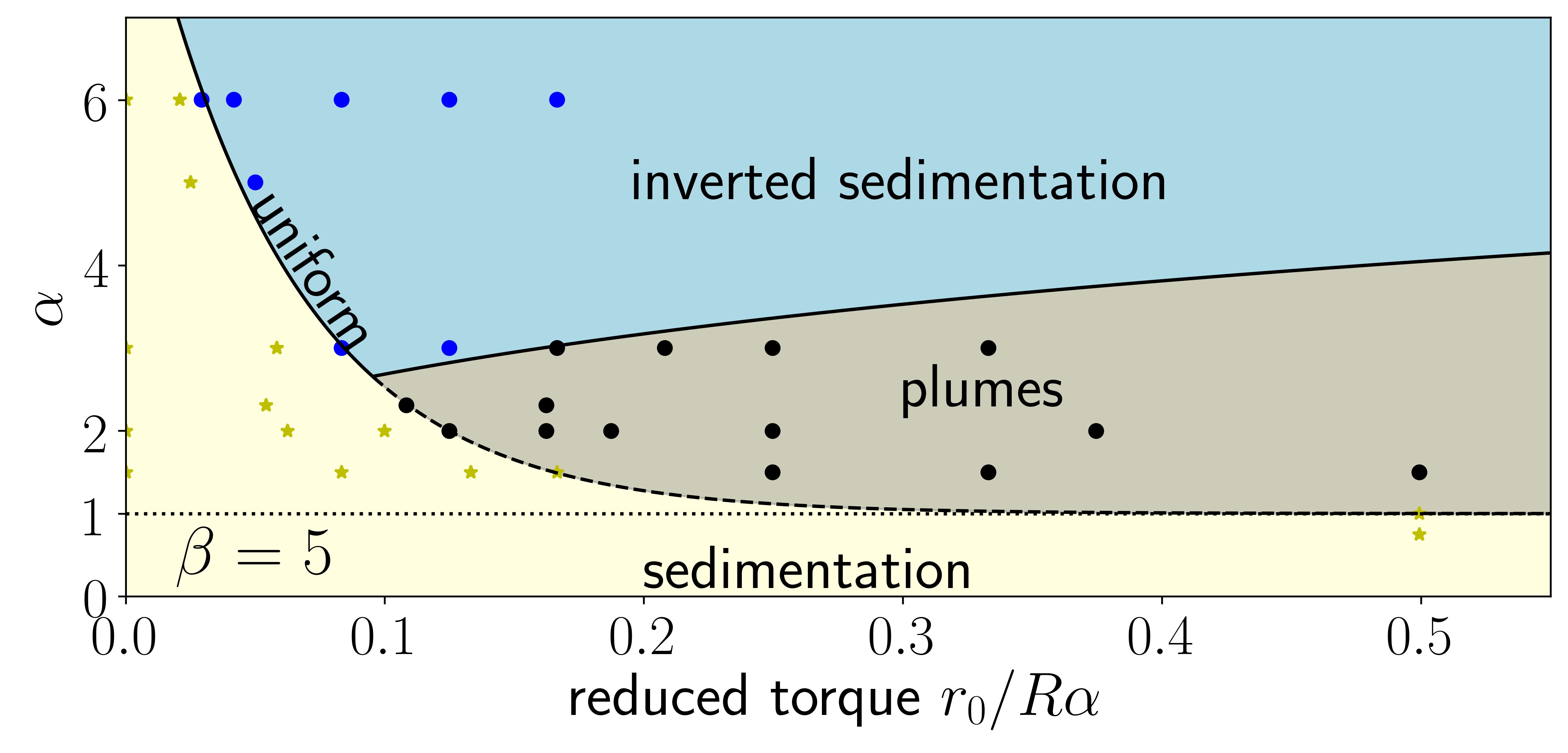}
\includegraphics[width=0.49\textwidth]{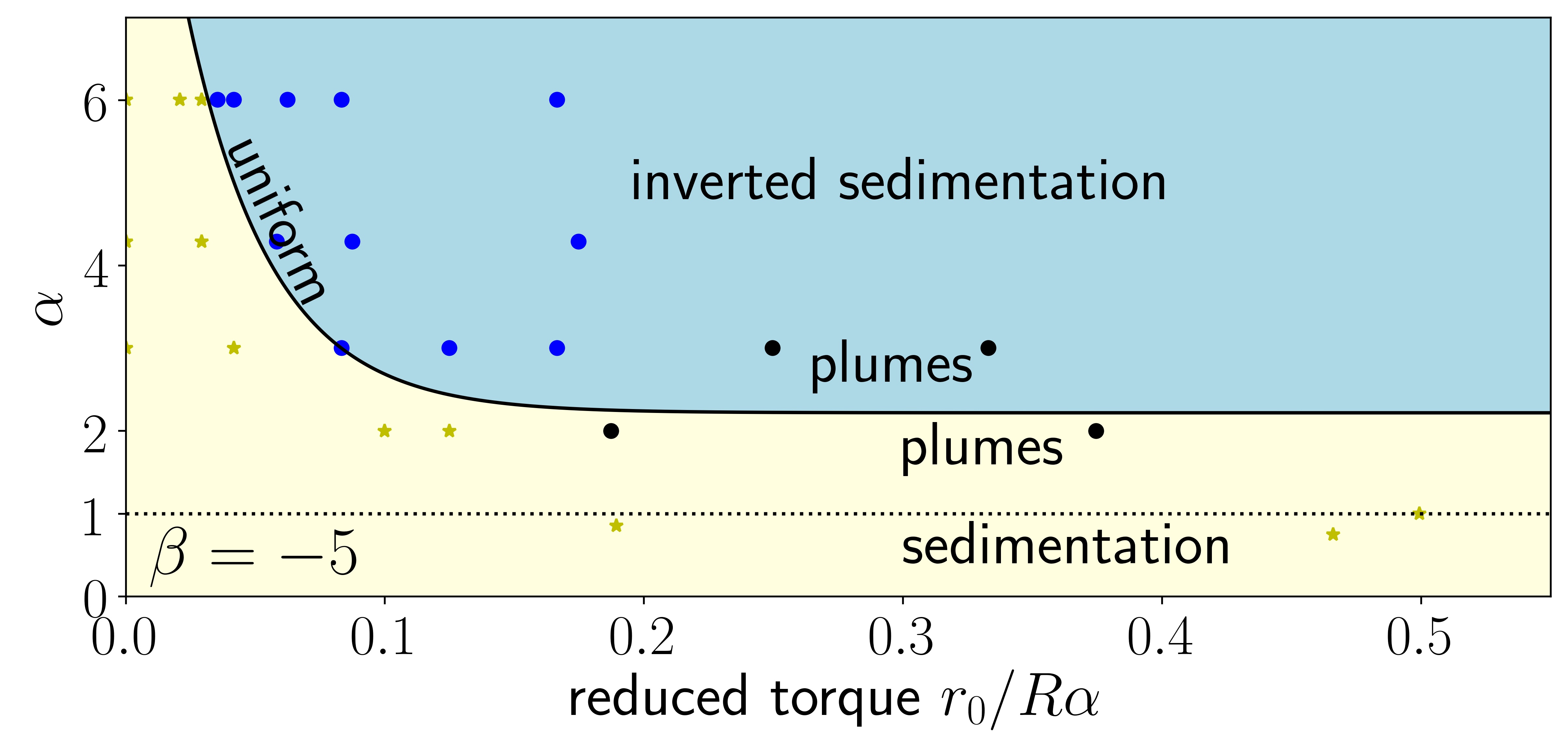}
\caption{Schematic state diagrams of strong puller (left, $\beta=5$) and pusher (right, $\beta=-5$) squirmers in the parameter space $\alpha$ versus $r_0/R\alpha$. 
The solid line separates the sedimenation from inverted sedimentation. For pullers the region for plume occurence is rationalized in sect.\ \ref{subsec.plumes_pullers_pushers}.
}
\label{fig:pusher_puller_states}
\end{figure*}

\subsection{Influence of squirmer type: pushers and pullers}
\label{sec:flow_fields}

For strong pullers and pushers the state diagrams, which we show in fig.\ \ref{fig:pusher_puller_states},
simplify considerably compared to neutral squirmers shown in fig.\ \ref{fig:neutral_states}(a). First of all, we do not observe any stable or transient convective rolls nor spawning clusters. The main features in the state diagrams of 
fig.\ \ref{fig:pusher_puller_states} are conventional and inverted sedimentation, where the separation line is shifted to higher 
torques and higher alpha compared to neutral squirmers. By visual inspection we observe plumes 
(see videos M11 and M12), where clusters of squirmers form in the upper region of the simulation cell, sink down, and dissolve. For pullers the plumes are  more pronounced as we explain in sect.\ \ref{subsec.plumes_pullers_pushers} and we indicate them by the grey shaded 
region as part of the sedimentation state.

\subsubsection{Influence of hydrodynamics on sedimentation state}

\begin{figure}
\centering
\includegraphics[width=\linewidth]{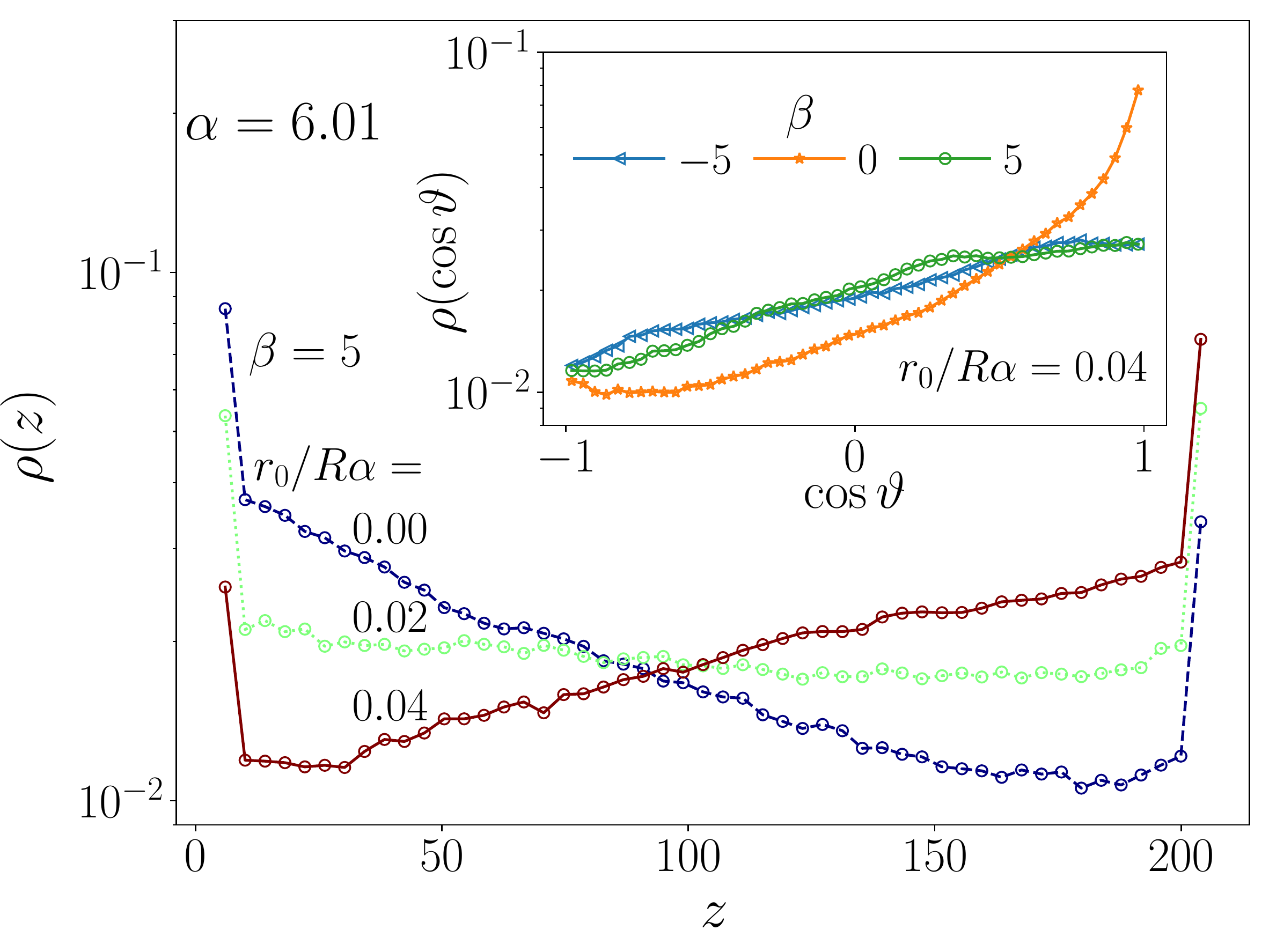}
\caption{Density profiles of strong pullers at $\alpha=6.01$ for different rescaled torques $r_0/R\alpha$, which generate exponential, constant and inverted exponential profiles. Inset: Orientational distribution function
at $\alpha=6.01$ and $r_0/R\alpha = 0.04$ for different squirmer parameters $\beta=-5,0,5$.}
\label{fig:pusher_puller_histograms}
\end{figure}

\begin{figure}
\centering
\includegraphics[width=\linewidth]{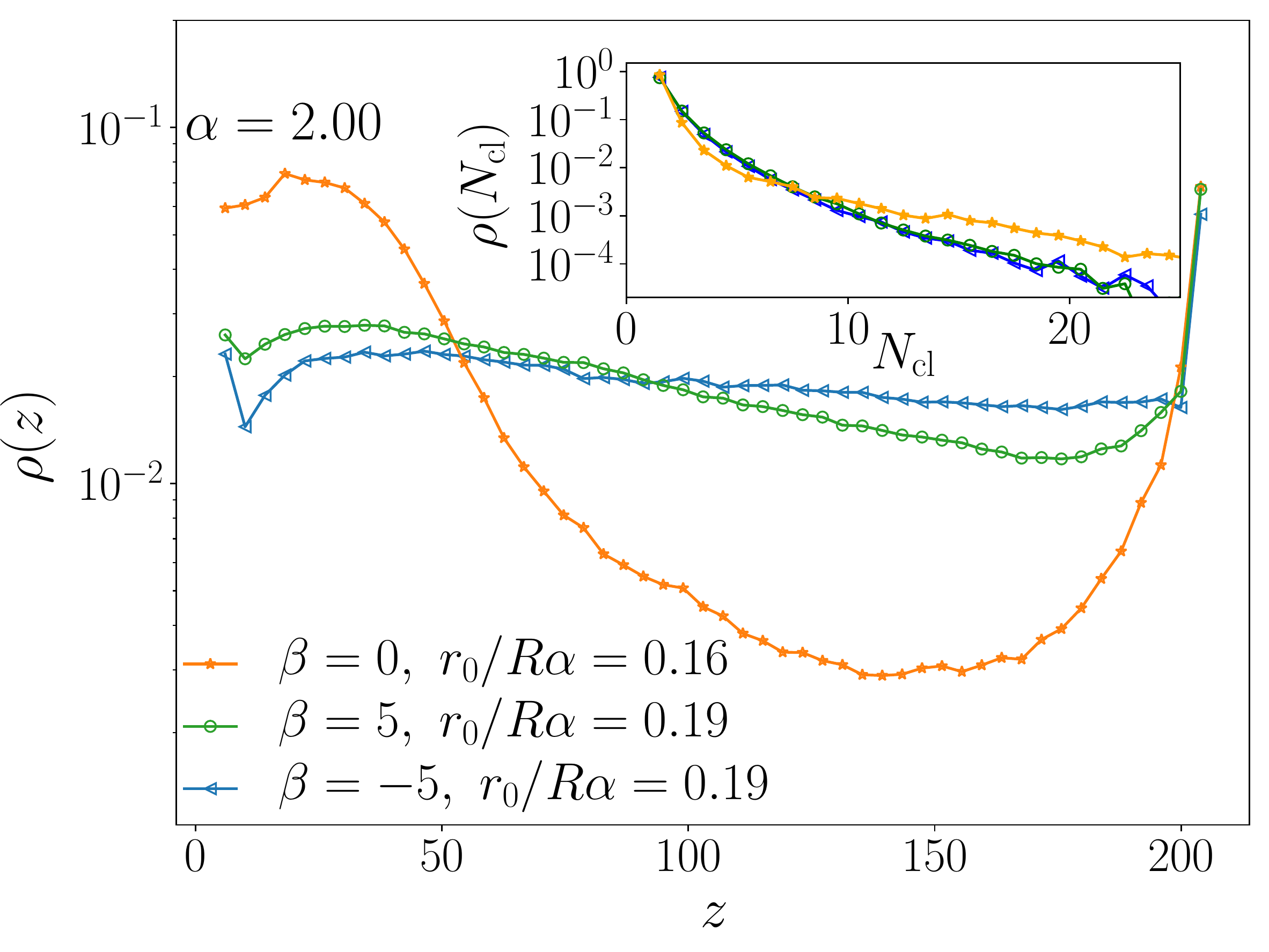}
\caption{Density profiles at $\alpha=2.00$ for neutral, pusher, and puller squirmers. The
rescaled torque $r_0/R\alpha$ is chosen such that collective sinking via
plumes occur in the system (see videos M11, M12 for $\beta=\pm 5$). 
Inset: Distribution of cluster sizes in the plumes for the same parameters as in the main plot.
} 
\label{fig:plumes_comparison}
\end{figure}

For pushers and pullers, which have a non-zero squirmer parameter $\beta$, the flow field of a force dipole is added to the total squirmer velocity field $\mathbf{u}(\mathbf{r})$ as documented by eq.\ (\ref{eq:squirmer_field}). 
It decays like $1/r^2$ and in contrast to the pure source-dipole field of neutral squirmers possesses a non-zero vorticity, which acts on the orientation of nearby squirmers. Previous studies already investigated the consequences of this vorticity
field for suspensions of microswimmers. They found that it weakens polar order in clusters, whereby pullers often retain a higher degree of polar order than pushers~\cite{HennesStark2014,EvansLauga2011,AlarconPagonabarraga2013,PessotMenzel2018}.

Indeed, in the inset of fig.\ \ref{fig:pusher_puller_histograms} we observe for the same parameters $\alpha = 6.01$ and 
$r_0/R\alpha =0.04$ that the neutral squirmer has a stronger alignment along the vertical than pushers and pullers. As a 
result, pushers and pullers have a stronger tendency to sink under gravity. Therefore, while neutral squirmers show inverted sedimentation at $\alpha = 6.01$ for all torque values, strong pullers ($\beta = 5$) exhibit a transition from conventional 
to inverted sedimentation with increasing $r_0/R\alpha$. Examples for the respective sedimentation profiles are plotted in the 
main graph of fig.\ \ref{fig:pusher_puller_histograms} together with a uniform profile right at the transition. Thus, the disturbance 
of the vertical alignment by the vorticity field of strong pushers and pullers shifts the transition line between conventional and
inverted sedimentation to larger torques and swimming velocities. For the same reason the tendency to form layers at the
upper wall is strongly reduced. This becomes obvious by comparing the sedimentation profiles for the same parameters
$\alpha = 6.01$ and $r_0/R\alpha =0.04$ for neutral squirmers in fig.\ \ref{fig:g-005_density_profiles} and strong pullers 
($\beta = 5$) in fig.\ \ref{fig:pusher_puller_histograms}.

\subsubsection{Plumes of pullers and pushers}
\label{subsec.plumes_pullers_pushers}

Unlike for neutral squirmers we do not observe stable convective rolls for both strong pullers and pushers. This becomes obvious from the density profiles in fig.~\ref{fig:plumes_comparison}. The broad and high density peak near the bottom wall is missing for pullers and pushers. Note that we increased the rescaled torque $r_0/R\alpha$ for pullers and pushers to be clearly in the region where plumes occur. Convective rolls possess dense squirmer clusters with some polar order due to bottom heaviness. 
However, such clusters are hydrodynamically unstable for squirmers with a strong force-dipole contribution
\cite{EvansLauga2011,AlarconPagonabarraga2013}. The same applies to spawning clusters, which we also do not observe.

\begin{figure}
\centering
\includegraphics[width=\linewidth]{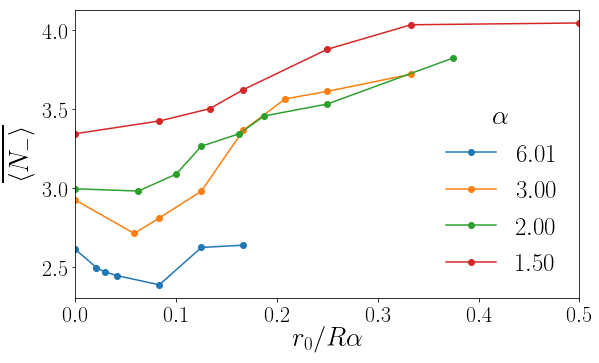}
\caption{Mean size of sinking puller clusters ($\beta=5$) as a function of the rescaled torque for different $\alpha$. The mean $\overline{\langle N_{\mathrm{-}}\rangle}$ is calculated in the lower region $20 \leq z/a_0 \leq 120$ and by averaging over more than $5\cdot 10^5$ time steps. Furthermore, all sinking clusters with $N_{\mathrm{-}} \ge 2$ are considered.}
\label{fig:mean_sinking_size}
\end{figure}

For neutral squirmers we observed plumes feeding 
convective rolls. For strong pullers and pushers we can also identify plumes by visual inspection, as demonstrated in videos M11 and M12, respectively. Pullers form visible plumes in the upper region that disband close to the bottom wall. Pusher plumes are very unstable 
and disintegrate already while they are sinking. Generally, the plume clusters are smaller compared to neutral squirmers. Thus, 
when plotting the mean number of sinking squirmers in a cluster, 
$\langle N_- \rangle$, versus time as in fig.\ \ref{fig:mean_sinking}, the plumes cannot clearly be identified by high spikes. 
Instead, for pullers we determined the mean squirmer number by also averaging over more than $5\cdot 10^5$ time steps.
The result is plotted in fig.\ \ref{fig:mean_sinking_size} \textit{versus} the rescaled torque for different swimming speeds $\alpha$.  
For $\alpha = 1.5$, 2.0, and 3.0 we roughly see a sigmoidal 
shape and locate a transition to a plume state at the inflection point. This is the meaning of the curved dashed line in the schematic state diagram of fig.\ \ref{fig:pusher_puller_states}, left.
For pushers we do not see such a sigmoidal shape but rather a slow and steady increase. For this reason we did not identify a separate region for plumes but only indicate that we see them around the line separating conventional and inverted sedimentation from each other at higher torques.

\section{Conclusions and outlook}
In this article we have investigated the remarkable features of microswimmer suspensions under gravity using full hydrodynamic simulations of ca. 900 squirmer model swimmers, where we concentrated on the neutral squirmer but also looked at strong pushers and pullers. We have determined the respective state diagrams varying the ratio of swimming to bulk sedimentation velocity and the gravitational torque due to bottom heaviness. The general trend in all three cases reveals conventional sedimentation for low swimming velocity and torque, while the sedimentation profile becomes inverted when increasing both values.

In addition, for neutral squirmers we have discovered a rich phenomenology in between both sedimentation states. Squirmers sink collectively in plumes due to reduced hydrodynamic friction and  feed fascinating convective roll patterns of elongated or toroidal shape that self-organize at the bottom of the system. The plume formation is supported by squirmer reorientation due to  vorticity resulting from the stokeslet contribution to the flow field. The latter is induced by the gravitational force acting on each squirmer. The combination of upward swimming (gravitaxis) and reorientation by nearby flow fields (rheotaxis) is called gyrotaxis, a mechanism introduced and discussed in connection with bioconvection \cite{PedleyKessler1992,HillKessler1989,BeesHill1997,GhoraiHill1999,DesaiArdekani2017}.
In our case plumes and convective rolls also form transiently at larger torques. When starting from an initially uniform squirmer distribution,
a dense layering of squirmers forms at the top wall. 
It develops an instability due to the gyrotactic mechanism and then sinking plumes emerge. At increasing torque and moderate speed ratios we also observe dense but porous squirmer clusters that float above the bottom wall and spawn single squirmers. They also become 
transient for increasing speed ratio.

For strong pushers and pullers the transition line between conventional and inverted sedimentation is shifted to higher torques and speed ratios. The reason is the non-zero vorticity of the additional force-dipole flow field, which reorients neighboring squirmer orientations from the vertical so that they can sink more easily. This is also the reason why strong pusher and pullers do not show such a rich phenomenology. Only weak plume formation is observed without any stable convective rolls occurring. In the case of pullers we could quantify it by the mean size of sinking puller clusters. 

In systems with biological microswimmers plume formation is often traced back to the gyrotactic mechanism introduced above \cite{PedleyKessler1992,BeesHill1997,GhoraiHill1999}. However, also the overturning instability of a dense layer of microswimmers at the top boundary, reminiscent of the Rayleigh-Taylor instability is discussed \cite{ChildressSpiegel1975,HarashimaFujishiro1988,MogamiBaba2004}.
Indeed, in ref.\ \cite{MogamiBaba2004} for plumes of the microorganism \textit{Tetrahymena} emanating from a dense top layer, 
gyrotaxis is discarded. In contrast, in our case plumes for the convective rolls also form in bulk, where clearly gyrotaxis is the relevant mechanism. Even for the transient plumes of neutral squirmers emanating from a dense squirmer layer at the top boundary, we think that gyrotaxis is predominant. 
However, these comments also suggest that gyrotaxis of squirmers needs to be studied further in the future, in particular, how the characteristic length scales of plumes and convective rolls depend on system size and squirmer density.

In biological systems, microswimmers are often of pu\-sher and puller type but nevertheless, in contrast to 
our simulations, can form stable plumes and convection cells \cite{CzirokKessler2000,JanosiHorvath1998,BeesHill1997}.
Our investigations are performed for strong pushers and pullers, while sufficiently weak
pusher and puller squirmers with $\beta$ closer to zero will also show convective rolls. Indeed, estimates for the force-dipole moment of some biological microswimmers show a weak dipole strength 
\cite{BerkeLauga2008,DrescherTuval2010,DrescherGoldstein2011}.
Furthermore, in our simulations we look at a very generic system concentrating on gravitational force, bottom heaviness, and hydrodynamic interactions between the squirmers. Real microswimmers with a non-spherical shape also experience a drag torque \cite{SenguptaStocker2017,Roberts2010} and their flagella might act on neighbors by steric forces.  Also, for the algae \textit{C.reinhardtii} it has been shown that the flow field induced by the periodic beating pattern varies in time and 
during a short period is reminiscent of a pusher\ \cite{MuellerThiffeault2017,MathijssenPolin2018}. Thus, real microorganisms show a large variability 
and it was argued that different organisms could even be distinguished via their bioconvection patterns~\cite{CzirokKessler2000}. Generalizing our simulations in these directions provides opportunities for interesting future research.

Bottom heaviness is not the only source for orientational order in order to observe interesting pattern formation under gravity. It can also be induced by additional external fields, in which the microorganism performs taxis. For example, the alga \textit{C.reinhardtii} 
relies on phototaxis \cite{SatoToyoshima2018,SinghFischer2018}, while the bacterium \emph{B. subtilis} shows aerotaxis, where 
it aligns along a gradient of oxygen \cite{CzirokKessler2000,JanosiHorvath1998}.
The response to chemical fields (chemotaxis) is fascinating on its own \cite{BergBrown1972,TheurkauffBocquet2012,SahaRamaswamy2014,PohlStark2014,Stark2018,StuermerStark2019} 
and combining it with microswimmers~\cite{HuangKapral2017} moving under gravity opens a new research direction.

Finally, microorganisms moving under gravity might also adapt their behavior to further external cues. For example, it has
been argued that phytoplankton actively change their morphology to adjust their migration strategy to turbulent flow fields \cite{SenguptaStocker2017}. This connects to another new and fascinating research direction related to learning in active systems \cite{ColabreseBiferale2017,Muinos-LandinCichos2018,SchneiderStark2019}.

\begin{acknowledgement}
We are grateful for stimulating discussions with and valuable input from J.-T. Kuhr, R. Kapral, K. Drescher, and A.J.T.M. Mathijsen, and thank the referees for their suggestions. 
This project was funded by Deut\-sche For\-schungs\-ge\-mein\-schaft through the priority program SPP1726 (grant number STA352/11). The authors acknowledge the North-German Supercomputing Alliance (HLRN) for providing HPC resources that have contributed to the research results reported in this paper.
\end{acknowledgement}

\section{Appendix}
Table \ref{tab:movies} provides an index of the videos referenced in this paper 
and made available in the electronic supplemental material. It contains the system parameters and describes the states visualized by the videos.

\begin{table*}
\centering
\caption{Index of videos provided in the supplemental material.}
\begin{tabular}{c|c|c|c|l}
movie name & $\beta$ & $\alpha$ & $r_0/R\alpha$ & state \\
\hline
M1 & $0$ & $6.01$ & $0.02$ & inverted sedimentation\\
M2 & $0$ & $1.50$ & $0.01$ & sedimentation \\
M3 & $0$ & $2.31$ & $0.11$ & plumes and convective rolls\\
M4 & $0$ & $1.50$ & $0.50$ & spawning cluster \\
M5 & $0$ & $2.31$ & $0.11$ & plumes and convective rolls (3D)\\
M6 & $0$ & $2.31$ & $0.11$ & plumes and convective rolls, larger system\\
M7 & $0$ & $2.31$ & $0.11$ & plumes and convective rolls, larger system (3D)\\
M8 & $0$ & $6.01$ & $0.08$ & evaporating plume \\
M9 & $0$ & $2.31$ & $0.22$ & transient roll\\
M10 & $0$ & $1.50$ & $0.17$ & spawning cluster\\
M11 & $5$ & $2.00$ & $0.19$ & weak plumes \\
M12 & $-5$ & $2.00$ & $0.19$ & weak plumes
\end{tabular}
\label{tab:movies}
\end{table*}

\section{Authors contributions}
All the authors were involved in the preparation of the manuscript. All the authors have read and approved the final manuscript.

\bibliographystyle{epj}
\bibliography{shorttitles,lit}

\end{document}